\documentclass{jfm} 
\usepackage{upmath,amssymb,amsbsy} 

\usepackage{graphicx,natbib,empheq,relsize}  
\usepackage{pstool} 
\usepackage[normalem]{ulem}

\providecommand\bnabla{\boldsymbol{\bld{\nabla}}}
\providecommand\bcdot{\boldsymbol{\cdot}}
 
\usepackage{mathrsfs} 
\usepackage{multirow}
\usepackage{amsmath,mathtools}
\usepackage[usenames,dvipsnames]{xcolor} 

\newcommand{\tr}{\textrm{tr}\,}
\newcommand{\sym}{\textrm{sym}\,} 
\newcommand{\asym}{\textrm{asym}\,} 
\newcommand{\Pos}{{\textbf{Pos}}}
\newcommand{\Sym}{{\textbf{Sym}}}
\newcommand{\GL}{{\textbf{GL}}}
\newcommand{\SO}{{\textbf{SO}}}
\newcommand{\dd}[2]{{ \frac{\text{D} {#1}}{\text{D} {#2}}   }}

\newcommand{\bld}[1]{{\boldsymbol{ #1 }} } 
\DeclareMathAlphabet{\mathscrbf}{OMS}{mdugm}{b}{n} 
\newcommand\Wie{\mbox{\textit{Wi}}}  
\newcommand{\e}[1]{{\text{e}^{ #1 }} } 

\def\sldsh{\rule[0.2\baselineskip]{0.075in}{0.65pt}}
\def\sldot{\rule[0.2\baselineskip]{0.025in}{0.65pt}}
\def\bldot{\hspace{0.025in}}
\def\linesolids{\rule[0.2\baselineskip]{0.275in}{0.65pt}}
\def\linedashed{\sldsh\bldot\sldsh\bldot\sldsh} 
\def\linedshdot{\sldsh\bldot\sldot\bldot\sldsh\bldot\sldot} 
\def\linedotted{\sldot\bldot\sldot\bldot\sldot\bldot\sldot\bldot\sldot\bldot\sldot}

\def\linesolidsSymTriangleUp{\rule[0.2\baselineskip]{0.1375in}{0.65pt}\parbox{0.11in}{$\vspace{0in}\hspace{-0.025in}\mathlarger{\bigtriangleup}$}$\hspace{-0.045in}$\rule[0.2\baselineskip]{0.1375in}{0.65pt}}
\def\linesolidsSymDiamond{\rule[0.225\baselineskip]{0.1375in}{0.65pt}\parbox{0.135in}{$\hspace{-0.0025in}\vspace{0in}\mathlarger{\mathlarger{\mathlarger{\mathlarger{\diamond}}}}$}$\hspace{-0.03675in}$\rule[0.225\baselineskip]{0.1375in}{0.65pt}}

\def\linesolidsSymCirc{\rule[0.225\baselineskip]{0.1375in}{0.65pt}\parbox{0.0825in}{{$\vspace{-0.01in}\hspace{-0.015in}\mathlarger{\mathlarger{\mathlarger{\mathlarger{\circ}}}}$}}\rule[0.225\baselineskip]{0.1375in}{0.65pt}}

\newcommand\redfilledcirc{$\mathlarger{\mathlarger{{\color{red}\bullet}\mathllap{\color{red}\circ}}}$}

\newcommand{\rev}[1]{{\color{blue}#1}}

\title[Perturbative expansions of the conformation tensor]{Perturbative expansions of the conformation tensor in viscoelastic flows}

\author[Hameduddin, Gayme \& Zaki]%
{Ismail Hameduddin,\ns 
	Dennice F.\,Gayme 
	and Tamer  A.\,Zaki \corresp{\email{t.zaki@jhu.edu}}
}

\affiliation{
	Department of Mechanical Engineering, The Johns Hopkins University,
	Baltimore, MD 21218, USA }

\pubyear{2018}
\volume{xxx}
\pagerange{xxx--xxx}
\date{\today; revised ?; accepted ?. - To be entered by editorial office}

\begin{document}
\maketitle

\begin{abstract} 
We consider the problem of formulating perturbative expansions of the conformation tensor, which is a positive--definite tensor representing polymer deformation in viscoelastic flows.
The classical approach does not explicitly take into account that the perturbed tensor must remain positive definite\textemdash a fact that has important physical implications, e.g.\,extensions and compressions are represented similarly to within a negative sign, when physically the former are unbounded and the latter are bounded from below.  
Mathematically, the classical approach assumes that the underlying geometry is Euclidean, and this assumption is not satisfied by the manifold of positive definite tensors.
We provide an alternative formulation that retains the conveniences of classical perturbation methods used for generating linear and weakly nonlinear expansions, but also provides a clear physical interpretation and a mathematical basis for analysis.
The approach is based on treating a perturbation as a sequence of successively smaller deformations of the polymer.
Each deformation is modelled explicitly using geodesics on the manifold of positive--definite tensors.
Using geodesics, and associated geodesic distances, is the natural way to model perturbations to positive--definite tensors because it is consistent with the manifold geometry.
Approximations of the geodesics can then be used to reduce the total deformation to a series expansion in the small perturbation limit.
We illustrate our approach using direct numerical simulations of the nonlinear evolution of Tollmien--Schlichting waves.
\end{abstract}

\section{Introduction}
In viscoelastic flows, the conformation tensor is a second-order, positive--definite tensor used to represent the polymer deformation.
When analysing such flows, we are frequently interested in generating small perturbations to a given base flow conformation tensor, e.g. for linear stability analysis or for deriving solutions for a flow that can be cast as a small perturbation of a known flow solution.
The standard approach is a generalization of the one used for the velocity field: the perturbed conformation tensor is expressed as the sum of the base flow tensor and a perturbation tensor.  
The latter is a symmetric tensor written as a series expansion in a small parameter.
Such an approach, generally referred to as a weakly nonlinear expansion, has been used in a myriad of different ways in the literature and has proven useful for extracting important mechanisms from the governing equations.
As opposed to the velocity field, however, this approach has several important limitations when applied to the conformation tensor, including: (a) the perturbation tensor does not describe material deformation, unlike the full conformation tensor, and does not have a known physical interpretation, (b) finite-amplitude perturbations are not possible in general and (c) the norm of the perturbation tensor is not an appropriate metric to quantify the size of the perturbation.
These limitations arise because the conformation tensor belongs to the set of positive--definite tensors whose properties must be taken into account when defining perturbations. 
In this paper, we tackle this problem by developing a framework to generate perturbations that are consistent with the physical interpretation of the conformation tensor.
Our framework helps us reconcile classical linear stability analysis and weakly nonlinear expansions with the physical interpretation of the conformation tensor and the geometry of the set of positive--definite tensors.

It is important to note that classical perturbation expansions have yielded a number of important predictions regarding viscoelastic flows that have been confirmed experimentally. 
We review some of these studies below.
An infinitesimally small linear perturbation is the simplest form of the perturbation tensor and has been widely used for linear stability analysis and also for energy amplification.
Early work on linear stability analysis predicted purely elastic two-dimensional instabilities in flows with curved streamlines \citep{Larson1990,Joo1992}, a result that was later confirmed by \citet{Groisman2000,Groisman2004} in low Reynolds number experiments.
Around the same time,  \citet{Avgousti1993} used bifurcation analysis in Taylor--Couette flow and demonstrated three-dimensional instabilities in the presence of elasticity.
\citet{McKinley1996} then developed a dimensionless criterion that predicts the critical conditions required for the onset of purely elastic instabilities in curved geometries.
These important discoveries increased interest in possible curvature-independent elastic instabilities, particularly in viscoelastic channel flow.
\citet{Jovanovic2011} and \citet{Page2014} found that linearly stable perturbations to the conformation tensor in this flow can amplify significantly due to transient growth via purely elastic mechanisms \citep[see][for an exposition on non-modal amplification due to operator non-normality]{Trefethen2005}. 
The most amplified non-modal disturbances are generally three-dimensional \citep{Hoda2008,Hoda2009}. 
The disturbance amplification, which exists even in the complete absence of inertia \citep{Jovanovic2010}, may be sufficient to trigger nonlinear instabilities. 
A purely elastic non-modal route to instability was also anticipated by \citet{Doering2006}.
\citet{Meulenbroek2004} used weakly nonlinear expansions  to show that sufficiently large transient growth in a viscoelastic channel flow acting over a slow time scale appears as a streamline curvature-inducing modification to the base-state, thereby producing the necessary conditions for a fast time scale curved streamline instability.
These theoretical results predicting an elastic instability in channel flow were experimentally confirmed by \citet{Pan2013} and \citet{Qin2017}, who found a turbulent-like state in a channel flow at low Reynolds number.
Perturbation expansions have also been used to revisit other important, classical flow problems in the presence of viscoelasticity. 
\citet{Page2015} predicted the existence of a the reverse Orr mechanism, whereby spanwise vorticity amplifies when aligned with the shear.  
More recently, \citet{Page2016} considered laminar viscoelastic channel flow with a wavy wall and used perturbation expansions to derive reduced dynamics in various asymptotic limits.
That work revealed the existence of an elastic critical layer that mediates the dynamics, and the theoretical predictions were partially reproduced in experiments by \citet{PhysRevFluids.3.091302}.

Despite the success described above, the standard approach to perturbing the conformation tensor is not fully satisfactory, as outlined below. 
Since the eigenvalues of the conformation tensor are the principal stretches of the polymer, the conformation tensor  is a positive--definite tensor.
Such tensors do not form a vector space.
As a result, several issues arise with the standard approach to perturbations that do not arise when we perturb a vector space quantity like the velocity field. 
Firstly, the perturbation tensor used in the standard approach is not a conformation tensor, but rather a symmetric tensor that can only be interpreted component-wise.
Thus, quantities that depend on the tensorial nature of the perturbation do not carry the desired physical interpretation, e.g.\,the eigenvalues are not necessarily positive and are no longer representative of the principal stretches of the polymer.
Secondly, the perturbation magnitude may be severely limited because such perturbations are guaranteed to be valid only in an infinitesimal sense and thus a finite-amplitude perturbation may violate the positive--definiteness requirement on the conformation tensor.
The amplitude of the perturbation must therefore be constrained or verified against the positive--definiteness condition.
An example of such an issue arises when generalizing the approach by \citet{Stuart1958} to viscoelastic flows.
That approach uses a base-state augmented with the associated linear modes at finite-amplitude to describe a saturated nonlinear flow state.
A third issue that arises is that there is no appropriate functional norm that can be used to quantify the magnitude of the conformation tensor and the perturbation.
For example, in the context of energy stability \citet{Doering2006} noted that the elastic energy was problematic because it was not strictly a metric and did not satisfy the triangle inequality.
In prior work, these issues, and others that arise because the conformation tensor is not a vector space quantity, are frequently concealed because the polymer stress is used instead of the conformation tensor. 
While the former does not strictly need to be positive--definite, it is positive--definite up to an additive constant for most models of interest.
Furthermore, the dynamics are usually expressed in terms of the conformation tensor rather than the polymer stress.

In a recent paper, \citet{Hameduddin2018a} established a theoretical foundation to quantify fluctuations of the conformation tensor in fully turbulent viscoelastic flows\textemdash see \citet{Graham2018} for a review of that work.
The present study utilizes that foundation and extends it to formulate a geometrically and physically consistent approach to generate small perturbations to the conformation tensor, analogous to a weakly nonlinear expansion of the velocity field.
Our approach relies on exploiting the interpretation of the conformation tensor as the left Cauchy-Green tensor associated with the polymer deformation and the geometry of the manifold of positive--definite tensors.
The perturbation is cast as a sequence of successively smaller deformations to the base-state.
When specialized to a single deformation, our approach reduces to the standard approach used for linear perturbations but now with an explicit underlying physical interpretation and also an inherent geometric structure derived from the manifold geometry.
Our framework provides new physical insights into polymer dynamics, and has implications for studies utilizing small perturbations to the conformation tensor since it resolves the outstanding issues with the standard approach that were highlighted in this introduction.
Namely, the framework generates perturbation tensors that are physically meaningful as left Cauchy-Green tensors representing the perturbation polymer deformation.
It also provides a way to generate finite-amplitude perturbations whose size can be quantified using the geometric structure of the set of positive--definite tensors.
In addition, explicit relationships can be found between the present and standard approaches used for generating small perturbations, both linear and weakly nonlinear.
These relationships enrich our understanding of the approaches used so far in the literature.

The present paper is organized as follows:
The necessary background from \citet{Hameduddin2018a} is provided in \S\ref{sec:geometry} and \ref{subsec:geometric} for the benefit of the reader, and in order to ensure that the present development is fully self-contained. 
In \S \ref{sec:geometry} we introduce the geometry of the set of positive--definite tensors, and the geometric decomposition of the conformation tensor in \S\ref{subsec:geometric}. 
The main theoretical results are developed in the remainder of \S \ref{sec:perturbExpansions}, and the evolution equations for the perturbations are derived in \S \ref{sec:perturbEvolve}.
Finally, in \S \ref{sec:DNS}, we illustrate our approach using direct numerical simulations of the nonlinear evolution of viscoelastic Tollmien-Schlichiting waves \citep{Lee2017}.
The conclusions of the paper are offered in section \S \ref{sec:conclusions}.

\section{Geometry}  

\label{sec:geometry} 
The set of second-order positive--definite tensors, $\Pos_3$,  does not form a vector space because it is not closed under arbitrary linear combinations.
The standard zero element and additive inverses, such as those used in the space of symmetric second-order tensors, $\Sym_3$, are also not part of $\Pos_3$. 
As a result, the Frobenius norm, which is the Euclidean norm in $\Sym_3$, is not a meaningful quantity for the elements of $\Pos_3$, and thus the Euclidean distance between tensors is also not meaningful.
Geometrically, Euclidean paths in $\Pos_3$ are not analogous to straight lines in $\mathbb{R}^3$ because they cannot be arbitrarily extended. Namely, for the path
\begin{align}
\mathsfbi{X}(r) = (1-r)\mathsfbi{A} + r\mathsfbi{B}, \label{EuclidPath}
\end{align} 
the tensor $\mathsfbi{X}(r)$ is guaranteed to be in $\Pos_3$ for any $\mathsfbi{A},\mathsfbi{B}\in \Pos_3$ only if $r\in[0,1]$.  
The inability to coherently define magnitudes, distances, and shortest paths makes analysing the dynamics of quantities  without an underlying geometric structure a difficult proposition.

The celebrated success of linear stability theory, which is founded on infinitesimal additive perturbations to the conformation tensor, owes itself to the tangent space structure of $\Pos_3$.  	An application of Weyl's theorem can be used to show that $\Sym_3$ is the local tangent space everywhere on $\Pos_3$. 	Thus, a sufficiently small additive perturbation by a symmetric tensor keeps the base-state conformation tensor, $\mathsfbi{\overline{C}}$, on $\Pos_3$.  
By assuming that perturbations are arbitrarily small, linear stability theory usually ignores the precise sense in which the perturbation, or distance between $\mathsfbi{\overline{C}}$ and the perturbed tensor, must be sufficiently small.
This distance, which may be important in comparing the effect of different linear modes or for generating finite-amplitude perturbations in numerical calculations, cannot be evaluated using Euclidean distances because $\Pos_3$ is non-Euclidean.
	For example, let $\mathsfbi{\overline{C}}=\mathsfbi{I}$ and the conformation tensor, $\mathsfbi{C}$, be given by
	\begin{align} 
	\mathsfbi{C}=\mathsfbi{\overline{C}} + \varepsilon\mathsfbi{I}=(1+\varepsilon)\mathsfbi{I} \label{linApprox}
	\end{align}
	 where $\varepsilon \in \mathbb{R}$ is a perturbation parameter. 	The positive--definiteness constraint is satisfied for all positive $\varepsilon$, but we require $|\varepsilon| < 1$ if $\varepsilon$ is negative.  
	 However, the Euclidean distance from  $\mathsfbi{\overline{C}}$  is the same for both positive and negative $\varepsilon$.

	 The asymmetry between positive and negative perturbations arises because the eigenvalues of the conformation tensor represent the principal stretches of the polymer normalized by the thermodynamic equilibrium stretch. Thus eigenvalues greater than $1$ represent stretches, and those less than $1$ represent compressions.
Thus, the compression converse to the stretch $(1+|\varepsilon|)\mathsfbi{I}$ is given by
	 \begin{align}
	 \left(\frac{1}{1+|\varepsilon|}\right)\mathsfbi{I} = \left(1 - |\varepsilon| + |\varepsilon|^2 - |\varepsilon|^3 + \hdots \right)\mathsfbi{I}
	 \end{align}
	which means that a negative $\varepsilon$ in (\ref{linApprox}) is equivalent to a physical contraction up to $\mathcal{O}(\varepsilon^2)$, an approximation that may be inadequate. It is not clear how to generalize the approach used for the simplified example presented here to more general cases.
	This discussion  highlights the importance of defining a consistent geometry on $\Pos_3$ that allows us to measure distances and define shortest paths. 
	In what follows, we define such a non-Euclidean geometry on $\Pos_3$.
Although $\Pos_3$ does not form a vector space, it does have a rich geometric structure.
In particular, $\Pos_3$ is a Cartan-Hadamard manifold: a simply-connected, geodesically complete Riemannian manifold with seminegative curvature \citep{Lang2001}.
By an abuse of notation, we will refer to both the set of positive--definite tensors, as well as the associated Riemannian manifold, as $\Pos_3$. We will describe some important aspects of this geometry below. We refer the interested reader to \citet{Hameduddin2018a} and \citet{Lang2001} for more complete descriptions.

The set, $\Pos_3$, is an open subset of $\mathbb{R}^{3 \times 3}$ and is therefore a manifold.
By invoking the Fr\'{e}chet derivative, it is easy to show that $\Sym_3$ is the tangent space at each point on the manifold.
The Riemannian structure on $\Pos_3$ equips the tangent space at each $\mathsfbi{Y}\in \Pos_3$ with a scalar product
\begin{align}
[\bld{\mathcal{A}},\bld{\mathcal{B}}]_{\mathsfbi{Y}} =  \tr (\mathsfbi{Y}^{-1} \bcdot \bld{\mathcal{A}} \bcdot \mathsfbi{Y}^{-1} \bcdot \bld{\mathcal{B}}), \label{localMetric}
\end{align}
where $\bld{\mathcal{A}},\bld{\mathcal{B}}\in \Sym_3$.
The collection of all the scalar products forms the Riemannian metric on $\Pos_3$.
The Riemannian metric can be used to define a local distance metric at each $\mathsfbi{A}\in \Pos_3$.
The length of a path on $\Pos_3$ can be calculated by patching together the local distance functions. 

The manifold, $\Pos_3$, is geodesically complete:   there is a unique, distance-minimizing curve on $\Pos_3$ between every $\mathsfbi{A},\mathsfbi{B}\in\Pos_3$, called the geodesic, which can also be arbitrarily extended. The geodesic is given by
\begin{align}
\mathsfbi{X}(r)=
\mathsfbi{A} \#_r \mathsfbi{B} = 
\mathsfbi{A}^{\frac{1}{2}} \bcdot
\left(
\mathsfbi{A}^{-\frac{1}{2}} \bcdot
\mathsfbi{B} \bcdot 
\mathsfbi{A}^{-\frac{1}{2}}
\right)^r \bcdot
\mathsfbi{A}^{\frac{1}{2}}, \qquad
r\in [0,1]. \label{geoPath}
\end{align}
The curve $\mathsfbi{X}(r)$ remains a geodesic on $\Pos_3$ for any $r \in [a,b] \subset \mathbb{R}$.
Geodesically complete curves on a Riemannian manifold are analogous to straight lines in Euclidean space.

The minimizing distance between $\mathsfbi{A},\mathsfbi{B}\in\Pos_3$, the geodesic distance, is given by 
\begin{align}
d(\mathsfbi{A},\mathsfbi{B}) = 
\sqrt{\tr \log^2(\mathsfbi{A}^{-\frac{1}{2}} \bcdot
\mathsfbi{B} \bcdot 
\mathsfbi{A}^{-\frac{1}{2}})}. \label{geodesicDefn}
\end{align}
where $\log$ here refers to the matrix logarithm.
By the Hopf-Rinow theorem, $\Pos_3$ is a complete metric space under the distance function $d(\cdot,\cdot)$, and thus the geodesic distance is a natural analog to the standard 2-norm in a Euclidean space.
The distance has other properties that accord well with our natural geometric intuitions. We outline some of these below.
\begin{enumerate}
\item The Euclidean norm in Euclidean space is invariant under translations (affine invariance). In an analogous manner, the distance metric is invariant under the action of the general linear group, $\GL_3$, 
\begin{align}
d(\mathsfbi{A},\mathsfbi{B})  = 
d(\mathsfbi{Y} \bcdot\mathsfbi{A}\bcdot \mathsfbi{Y}^{\mathsf{T}} ,\mathsfbi{Y} \bcdot \mathsfbi{B} \bcdot \mathsfbi{Y}^{\mathsf{T}} ),
\end{align}
for any $\mathsfbi{Y} \in \GL_3$.

\item The distance traversed along the Euclidean path (\ref{EuclidPath}) is given by $|r|\| \mathsfbi{B}-\mathsfbi{A}  \|_{F}$, where $\| \cdot \|_{F}$ is the Frobenius norm and we restrict $r \in [0,1]$ in order to remain within $\Pos_3$.
Similarly, the distance along the geodesic (\ref{geoPath}) is given by $|r|\,d(\mathsfbi{A},\mathsfbi{B})$, but the use of a consistent geometry means we can let $r \in \mathbb{R}$.

\item The Euclidean distance between $\mathsfbi{A}$ and $\mathsfbi{B}$ is the same as the distance between $-\mathsfbi{A}$ and $-\mathsfbi{B}$. Similarly, we have for the geodesic distance
\begin{align}
d(\mathsfbi{A},\mathsfbi{B})  = 
d(\mathsfbi{A}^{-1},\mathsfbi{B}^{-1}).
\end{align}
This property is especially attractive from a physical point of view, since it means the distance metric treats expansions and compressions on an equal footing, unlike the Euclidean metric.
\end{enumerate}

The distance metric may be viewed as the Frobenius norm or square-root of the second moment-invariant of $\bld{\mathcal{B}}\equiv\log(\mathsfbi{A}^{-\frac{1}{2}} \bcdot \mathsfbi{B} \bcdot \mathsfbi{A}^{-\frac{1}{2}}) \in \Sym_3$.
This tensor is a tangent direction on $\Pos_3$. It can be shown \citep{Hameduddin2018a} that the first invariant of this tensor is given by
\begin{align} 
 \tr \bld{\mathcal{B}} = \log \left( \det \mathsfbi{B}/\det \mathsfbi{A} \right), \label{logVol}
\end{align} 
which means that $\tr \bld{\mathcal{B}} $ can be related to the ratio of the volumes (determinants) of the deformation ellipsoids associated with $\mathsfbi{A}$ and $\mathsfbi{B}$:
 $\tr \bld{\mathcal{B}}  < 0$ if the deformation $\mathsfbi{A} \rightarrow \mathsfbi{B}$ is compressive and  $\tr \bld{\mathcal{B}}  > 0$ if it is expansive.

In the next section, we exploit the geometric structure of $\Pos_3$ introduced in this section to develop an analogue of the weakly nonlinear expansions of vector space quantities, and specialize it to the case of linear perturbations.

\section{Perturbative expansions of the conformation tensor}
\label{sec:perturbExpansions}
\subsection{Geometric decomposition}
\label{subsec:geometric}

In order to generate perturbative expansions of the conformation tensor $\mathsfbi{C}$ about the base state $\mathsfbi{\overline{C}}$, we first define an appropriate fluctuating conformation. For this, we follow the approach by \citet{Hameduddin2018a}, which we outline below. Later, we will also exploit this approach to generate the perturbative expansion we are seeking.

\begin{figure}
	\centering
	\includegraphics[scale=1.0]{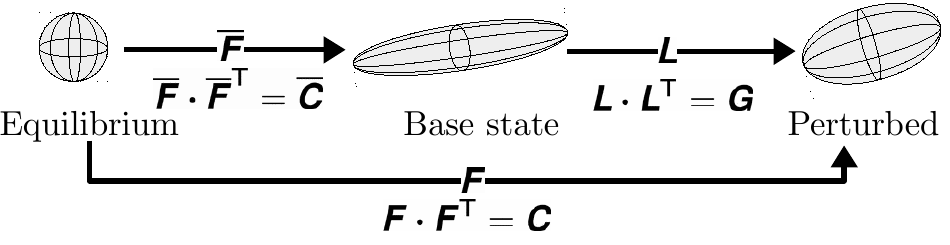}
	\caption{Schematic of the geometric decomposition, adapted from \citet{Hameduddin2018a}.}
	\label{fig:deformation_map}
\end{figure}

The conformation tensor, $\mathsfbi{C}$, is the left Cauchy-Green tensor associated with the polymer deformation \citep{Rajagopal2000,Cioranescu2016}
\begin{align}
\mathsfbi{C} = \mathsfbi{F}\bcdot \mathsfbi{F}^{\mathsf{T}}.
\end{align}
where $\mathsfbi{F}$ is the total deformation gradient that describes deformation with respect to the thermodynamic equilibrium.
We decompose this deformation into two: a deformation about the thermodynamic equilibrium that yields the base state, and a deformation about the base state that yields the total deformation. 
Accordingly, we decompose the deformation gradient as follows
\begin{align}
\mathsfbi{F} = \mathsfbi{\overline{F}} \bcdot \mathsfbi{L}
\end{align}
where $\mathsfbi{L}$ is the fluctuating deformation gradient, and $\mathsfbi{\overline{F}}$ is the deformation gradient associated with $\mathsfbi{\overline{C}}$,
\begin{align}
\mathsfbi{\overline{F}} = \mathsfbi{\overline{C}}^{\frac{1}{2}} \bcdot \mathsfbi{R}.
\label{Fbar}
\end{align}
Here $\mathsfbi{R} \in \SO_3$ and $\SO_3$ is the special orthogonal group.
 It is readily verified that $\mathsfbi{\overline{C}} = \mathsfbi{\overline{F}}  \bcdot \mathsfbi{\overline{F}}^{\mathsf{T}}$. 
In practice, we set $\mathsfbi{R}=\mathsfbi{I}$ (for example in \S\ref{sec:DNS}).
The fluctuating deformation gradient has the associated tensor $\mathsfbi{G} = \mathsfbi{L} \bcdot \mathsfbi{L}^{\mathsfbi{T}}$, which satisfies the relationship
\begin{align}
\mathsfbi{C} = \mathsfbi{\overline{F}}\bcdot\mathsfbi{G}\bcdot \mathsfbi{\overline{F}}^{\mathsf{T}}. \label{NecessaryDecomposition}
\end{align}
The tensor $\mathsfbi{G}$ is positive--definite and is equivalent to $\mathsfbi{C}$ but is transformed so that $\mathsfbi{C} = \mathsfbi{\overline{C}}$ if and only if $\mathsfbi{G}=\mathsfbi{I}$.
Thus $\mathsfbi{G}$ acts as a conformation tensor representing the fluctuation of $\mathsfbi{C}$ around $\mathsfbi{\overline{C}}$. 

The geometry of $\Pos_3$ can be used to quantify the fluctuating polymer deformation. 
Using (\ref{geodesicDefn}), the geodesic distance between $\mathsfbi{C}$ and $\mathsfbi{\overline{C}}$ can be written as
\begin{align}
 d(\mathsfbi{\overline{C}},\mathsfbi{C})=d(\mathsfbi{I},\mathsfbi{G})  = \sqrt{\tr \bld{\mathcal{G}}^2} 
\end{align}
where $\bld{\mathcal{G}} \equiv \log\mathsfbi{G}$, and is a measure of the mangnitude of the fluctuation. Similarly, we can evaluate whether a deformation is compressive or expansive with respect to the base state by examining the logarithmic volume ratio, $
 \tr \bld{\mathcal{G}}$.

\subsection{Weakly nonlinear deformations}
\label{sec:wnd}
\begin{figure}
	\centering
	\includegraphics[scale=1.0]{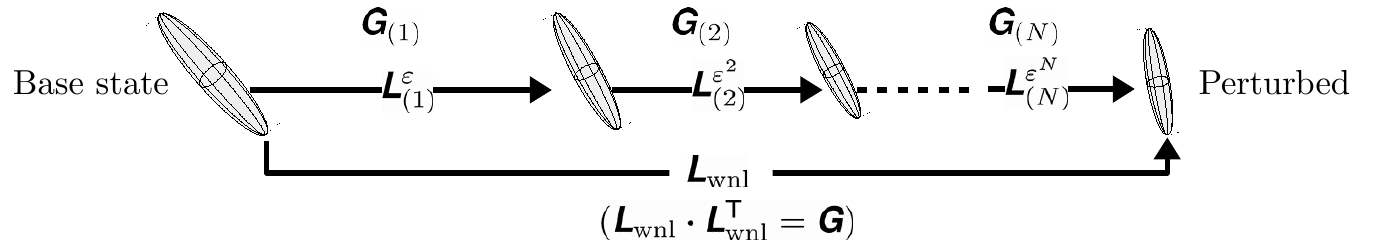}
	\caption{Schematic of a weakly nonlinear deformation, consisting of a sequence of successively smaller deformations.}
	
	\label{fig:deformation_map_series}
\end{figure}

A weakly nonlinear expansion up to the $N$-th power of the velocity field is given by
\begin{align}
\bld{u} &= \bld{\overline{u}} + \bld{u}' = \bld{\overline{u}} + \sum_{k=1}^{N}\varepsilon^{k} \bld{u}_{(k)} \label{uexpansion}
\end{align}
where $\bld{u}_{(k)}(\bld{x},t)$ for $k\in [1,N]$ are velocities.
A similar expansion for $\mathsfbi{C}$ is inappropriate because it is positive--definite and there is no \emph{a priori} guarantee on this property.
In order to obtain an analogous expansion for the conformation tensor, we generalize the approach we used in the previous subsection by multiplicatively decomposing the fluctuating deformation gradient into $N$ separate components. 
The construction of this fluctuating deformation gradient, denoted $\mathsfbi{L}_{\text{wnl}}$, through a series of successively smaller deformations is illustrated in figure \ref{fig:deformation_map_series}. Mathematically, we write
\begin{align}
\mathsfbi{L}_{\text{wnl}} = 
\mathsfbi{L}_{(1)}^{\varepsilon} \bcdot
\mathsfbi{L}_{(2)}^{\varepsilon^2} \bcdot
\hdots  \bcdot
\mathsfbi{L}_{(N)}^{\varepsilon^N}. \label{Ldecomp}
\end{align} 
We further assume that each $\mathsfbi{L}_{(k)}^{\varepsilon^k}$ in (\ref{Ldecomp}) is rotation-free.
By the polar decomposition and the requirement that $\det \mathsfbi{L}_{(k)} > 0$, this assumption implies that each $\mathsfbi{L}_{(k)}$ is positive--definite. 
Although each $\mathsfbi{L}_{(k)}$ is rotation-free, the overall fluctuating deformation gradient $\mathsfbi{L}_{\text{wnl}}$ given by (\ref{Ldecomp}) is not because the product of positive--definite tensors is not necessarily positive--definite.
The rotation appears if the principal axes of $\mathsfbi{L}_{(k)}$ and $\mathsfbi{L}_{(k')}$, when $k\neq k'$, are misaligned. 
The deformation gradient $\mathsfbi{L}_{\text{wnl}}$ is also not necessarily the same as $\mathsfbi{L}$ defined previously.
However, since $\mathsfbi{L}_{\text{wnl}}\bcdot \mathsfbi{L}_{\text{wnl}}^{\mathsf{T}}=\mathsfbi{L} \bcdot \mathsfbi{L}^{\mathsf{T}}=\mathsfbi{G}$, the polar decomposition can be used to show that $\mathsfbi{L}_{\text{wnl}}=\mathsfbi{V}\bcdot\mathsfbi{L}$ for some rotation tensor $\mathsfbi{V}$.

Each deformation gradient $\mathsfbi{L}_{(k)}^{\varepsilon^k}$ in (\ref{Ldecomp}) has an associated left Cauchy-Green tensor, $\mathsfbi{G}_{(k)}^{\varepsilon^{k}} = \mathsfbi{L}_{(k)}^{\varepsilon^k}\bcdot (\mathsfbi{L}_{(k)}^{\varepsilon^k})^{\mathsf{T}}$, which can be viewed as a geodesic of length $|\varepsilon|^k \| \bld{\mathcal{G}}_{(k)} \| \sim |\varepsilon|^k$ on $\Pos_3$ emanating from $\mathsfbi{I}$ and can be expressed conveniently as 
\begin{align}
\mathsfbi{G}_{(k)}^{\varepsilon^{k}} = \e{ \varepsilon^{k} \bld{\mathcal{G}}_{(k)}}  \label{Gk_defn}
\end{align}
where $\e{\bld{\mathcal{A}}}$ is the matrix exponential  of $\bld{\mathcal{A}}$, $\bld{\mathcal{G}}_{(k)} \in \Sym_3$ are tangents on $\Pos_3$, and $\bld{\mathcal{G}}_{(0)} = \bld{0}$.
With (\ref{Gk_defn}), it is easy to show that $\det  \mathsfbi{L}_{\text{wnl}} > 0$, which means that $\mathsfbi{L}_{\text{wnl}}$ is a physically admissible deformation gradient.

The tangents on $\Pos_3$, $\bld{\mathcal{G}}_{(k)}$, can be used to physically characterize the perturbation deformation.
The associated deformation gradient is given by $\mathsfbi{L}_{(k)}^{\varepsilon^k}= \e{\varepsilon^k\bld{\mathcal{G}}_{(k)}/2}$.
When $\bld{\mathcal{G}}_{(k)}$ is diagonal, $\mathsfbi{L}_{(k)}^{\varepsilon^k}$ is diagonal and thus the deformation is a shear-free distortion, or solely altering the volume ratio.
On the other hand, when $\tr \bld{\mathcal{G}}_{(k)} = 0$, then $\det\mathsfbi{L}_{(k)}^{\varepsilon^k}=1$, and the deformation is purely shearing, or volume-ratio preserving.
 
Using the square root of \eqref{Gk_defn} in \eqref{Ldecomp}, the left Cauchy Green tensor $\mathsfbi{G}$ is given by,
\begin{align}
\mathsfbi{G} &=  
\e{ \varepsilon \bld{\mathcal{G}}_{(1)}/2} \bcdot
\hdots \bcdot
\e{ \varepsilon^{N-1} \bld{\mathcal{G}}_{(N-1)}/2} \bcdot
\e{ \varepsilon^{N} \bld{\mathcal{G}}_{(N)}} 
\bcdot \e{ \varepsilon^{N-1} \bld{\mathcal{G}}_{(N-1)}/2} 
\bcdot \hdots 
\bcdot \e{ \varepsilon \bld{\mathcal{G}}_{(1)}/2} \label{nl_expansion0} \\
&= \mathsfbi{I} + \varepsilon  \bld{\mathcal{G}}_{(1)} +  \varepsilon^2 \left( \frac{\bld{\mathcal{G}}_{(1)}^2}{2} +   \bld{\mathcal{G}}_{(2)}\right) + 
\varepsilon^3 \left(\frac{\bld{\mathcal{G}}_{(1)}^3}{6}
+  \sym (\bld{\mathcal{G}}_{(1)}\bcdot\bld{\mathcal{G}}_{(2)})  + \bld{\mathcal{G}}_{(3)}\right) +
\hdots \label{nl_expansion1}
\end{align} 
The second equality made use of the matrix exponential $\e{ \varepsilon^{k} \bld{\mathcal{G}}_{(k)} } = \sum_{p=0}^{\infty} \varepsilon^{kp} \bld{\mathcal{G}}_{(k)}^p/p!$. 
The form (\ref{nl_expansion1}) is a series expansion of the conformation tensor that serves as an analogue to the weakly nonlinear expansion of the velocity in (\ref{uexpansion}). 
In fact, the terms in (\ref{nl_expansion1}) can be related to the standard weakly nonlinear expansion of the conformation tensor, $\mathsfbi{C} = \mathsfbi{\overline{C}} + \sum_{k=1}^{\infty} \varepsilon^k\mathsfbi{C}_{(k)}$. 
The difference between $\mathsfbi{C}_{(k)}$ and $\bld{\mathcal{G}}_{(k)}$ is that the latter can be related to a polymer perturbation deformation by means of the framework introduced above.
Furthermore, (\ref{nl_expansion1}) shows that the $\mathsfbi{C}_{(k)}$ are not independent of one another, even before the expansion is applied in the governing equations to examine the dynamics. 
This behaviour is consistent with the curved geometry of $\Pos_3$ because, unlike in Euclidean space, the characteristics of a particular perturbation depends on the location on the manifold where the perturbation is applied. 
In this view of the geometry, the deformation associated with $\mathsfbi{L}_{(n)}^{\varepsilon^{n}}$ is a perturbation to the deformation associated with $\mathsfbi{L}_{(1)}^{\varepsilon} \bcdot
\mathsfbi{L}_{(2)}^{\varepsilon^2} \bcdot \hdots  \bcdot \mathsfbi{L}_{(n-1)}^{\varepsilon^{n-1}}$ and thus the $n$-th order term in the series expansion must depend on all $\bld{\mathcal{G}}_{(k)}$ with $k=1,\hdots n$.
The behaviour is also consistent with a physical understanding of successive deformations of the polymer; a deformation is only sensible with respect to an existing configuration and is thus dependent on it from the point of view of an independent observer.

\begin{figure}
	\centering
	\includegraphics[scale=1.0]{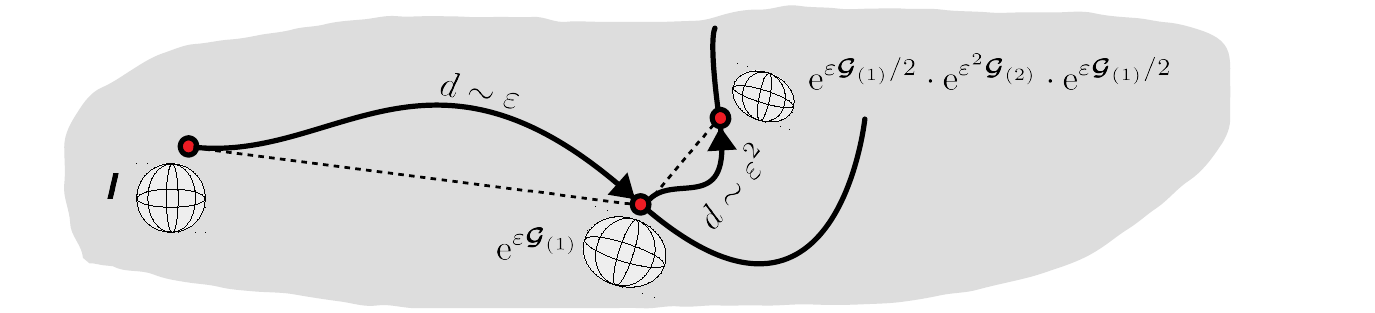}
	\caption{Illustration of weakly nonlinear deformation when $N=2$. In this case, the deformation corresponds to a piece-wise geodesic on $\Pos_3$. The thick black lines represent geodesics and dashed lines are Euclidean paths on $\Pos_3$.}
	\label{fig:connected_geodesic}
\end{figure}

One aspect of the relationship between the present approach of decomposing the total deformation into a series of successive deformations (cf. figure \ref{fig:deformation_map_series}) and the geometry of $\Pos_3$ is that the left Cauchy-Green tensor associated with each deformation is chosen to be a geodesic emanating from $\mathsfbi{I}$.
Another direct connection between geodesics on $\Pos_3$ and the overall deformation represented  by $\mathsfbi{G}$ can be made when $N\leq 2$. 
When $N=1$, $\mathsfbi{G}=\e{ \varepsilon \bld{\mathcal{G}}_{(1)}}$ is simply a geodesic of length $\sim \varepsilon$ emanating from $\mathsfbi{I}$ in the direction $\bld{\mathcal{G}}_{(1)}$. When $N=2$, we have
\begin{align}
\mathsfbi{G} &=   
\e{ \varepsilon  \bld{\mathcal{G}}_{(1)}/2} \bcdot
\e{ \varepsilon^{2} \bld{\mathcal{G}}_{(2)}} 
\bcdot \e{ \varepsilon  \bld{\mathcal{G}}_{(1)}/2},   
\label{nl_expansion1_2}
\end{align}
which implies that $\mathsfbi{G}$ is `piece-wise geodesic': it consists of a geodesic of length $\sim \varepsilon$ emanating from $\mathsfbi{I}$ in the direction $\bld{\mathcal{G}}_{(1)}$, followed by a geodesic of length $\sim \varepsilon^2$ in the direction $\bld{\mathcal{G}}_{(2)}$, as illustrated in figure \ref{fig:connected_geodesic}.
Such an interpretation is not generally possible for $N>2$ because then $\e{ \varepsilon \bld{\mathcal{G}}_{(1)}/2} \bcdot \hdots \bcdot \e{ \varepsilon^{N-1} \bld{\mathcal{G}}_{(N-1)}/2}$ need not be in $\Pos_3$.
If we assume that $\bld{\mathcal{G}}_{(1)}, \hdots, \bld{\mathcal{G}}_{(N-1)}$ are commutative with respect to multiplication, then 
\begin{align}
\e{ \varepsilon \bld{\mathcal{G}}_{(1)}/2} \bcdot \hdots \bcdot \e{ \varepsilon^{N-1} \bld{\mathcal{G}}_{(N-1)}/2} = \e{\frac{1}{2}\sum_{k=1}^{N-1} \varepsilon^k \bld{\mathcal{G}}_{(k)}} \in \Pos_3.
\end{align}
Thus, in this case, the interpretation of the successive deformations as a piece-wise geodesic on $\Pos_3$ holds for arbitrary $N$.
The inability to extend the piece-wise geodesic intepretation to arbitrary $N$ arises because successive deformations are, in general, physically misaligned and thus, by the polar decomposition, a rotation is imparted to the deformation gradient that depends on the order in which successive intermediate deformations were performed.
This order of the intermediate deformations is only relevant when $N\geq2$.
When the deformations are physically aligned, the overall deformation gradient is   positive--definite: it has no associated rotation and the order of the intermediate deformations can be arbitrarily changed.

An alternative to the present approach for generating a series expansion of $\mathsfbi{G}$ is to expand the tangent vector on $\Pos_3$, 
\begin{align}
\mathsfbi{G} = \exp\left(\sum_{k=1}^{N} \varepsilon^k \bld{\mathcal{G}}_{(k)}\right) = 
\mathsfbi{I} + \sum_{k=1}^{N} \varepsilon^k \bld{\mathcal{G}}_{(k)} + \frac{1}{2}\left(\sum_{k=1}^{N} \varepsilon^k \bld{\mathcal{G}}_{(k)}\right)^2 +\hdots \label{tanExp}
\end{align}
It is easy to show that our proposed approach (\ref{nl_expansion1}) and the alternative  (\ref{tanExp}) are  equivalent if   $\bld{\mathcal{G}}_{(1)}, \hdots, \bld{\mathcal{G}}_{(N)}$ are commutative with respect to multiplication.
When commutativity is not satisfied, the two approaches are still equivalent up to $\mathcal{O}(\varepsilon^4)$.
The drawback with using (\ref{tanExp}) is that the individual terms of the expansion cannot be associated with a polymer deformation because $\e{\mathsfbi{A} + \mathsfbi{B}} \neq \e{\mathsfbi{A}} \bcdot \e{ \mathsfbi{B}}$.
The expansion (\ref{tanExp}) also cannot be related to the geometry of $\Pos_3$ in the same way as (\ref{nl_expansion1}).

We developed an approach to generate a perturbation deformation of the polymers with arbitrarily many deformations. 
The magnitudes of the deformations are of successively higher order with respect to the distance metric on the manifold.
We next consider the case of linear perturbations, where a single deformation is involved.

\subsection{Linear perturbations}

We can generate a small perturbation about $\mathsfbi{\overline{C}}$ by translating the conformation tensor along a geodesic emanating from $\mathsfbi{\overline{C}}$. This can be accomplished by setting
\begin{align}
\mathsfbi{G} &= \mathsfbi{I} \#_{\varepsilon}  \e{\bld{\mathcal{G}}_{(1)}}  = \e{\varepsilon\bld{\mathcal{G}}_{(1)}} 
\label{ItoGPath} 
\end{align}
where $\bld{\mathcal{G}}_{(1)}$ is a prescribed tangent on $\Pos_3$, $\varepsilon \in \mathbb{R}$ and $d(\mathsfbi{I},\e{\varepsilon\bld{\mathcal{G}}_{(1)}}) = |\varepsilon| \| \bld{\mathcal{G}}_{(1)} \| \sim \varepsilon$. 
The parameter $\varepsilon$ here represents the amount of volumetric deformation encoded in the perturbation because the volume of $\mathsfbi{G}$ is given by $\det \mathsfbi{G} = \e{\varepsilon\tr\bld{\mathcal{G}}_{(1)}}$, or equivalently $\det \mathsfbi{C} = c^{\varepsilon}\det \mathsfbi{\overline{C}}$ for constant $c=\e{\tr\bld{\mathcal{G}}_{(1)}}$.
The expression (\ref{ItoGPath}) is valid, in a kinematic sense, for all $\varepsilon \in \mathbb{R}$. 
In the case when $\varepsilon \rightarrow 0$, we may approximate the matrix exponential, $ \e{\varepsilon\bld{\mathcal{G}}_{(1)}}= \sum_{k=0}^{\infty}\varepsilon^k\bld{\mathcal{G}}_{(1)}^k/k!$, as  
\begin{align}
\mathsfbi{G}= \e{\varepsilon\bld{\mathcal{G}}_{(1)}}= 
\mathsfbi{I} + \varepsilon  \bld{\mathcal{G}}_{(1)}   + \mathcal{O}(\varepsilon^2 \e{\varepsilon   })  \label{GTrunc1}
\end{align}
where  the truncation error is based on the bounds derived by \citet{Suzuki1976}. 
The result (\ref{GTrunc1}) is the same as the weakly nonlinear expansion (\ref{nl_expansion1}) with 
\begin{align}
\bld{\mathcal{G}}_{(2)} =  \bld{\mathcal{G}}_{(3)} = \hdots = \bld{\mathcal{G}}_{(N)} = \bld{0}.
\end{align}

Multiplying (\ref{GTrunc1}) by $\mathsfbi{\overline{F}}$ on both sides, we obtain
\begin{align}
\mathsfbi{C}= \mathsfbi{\overline{C}} + \varepsilon \mathsfbi{C}_{(1)} + \mathcal{O}(\varepsilon^2 \e{\varepsilon   }) \label{GTrunc2}
\end{align}
where  $\mathsfbi{C}_{(1)} = \mathsfbi{\overline{F}} \bcdot \bld{\mathcal{G}}_{(1)} \bcdot \mathsfbi{\overline{F}}^{\mathsf{T}}$, which is similar to the standard approach involving an additive perturbation to $\mathsfbi{\overline{C}}$. 
However, now the fluctuation, $\mathsfbi{C}'=\varepsilon\mathsfbi{C}_{(1)}$, has a clear interpretation as a tangent to the manifold at the base-point $\mathsfbi{\overline{C}}$. 
Furthermore, the normalization of $\mathsfbi{C}'$  is proportional to the geodesic distance away from $\mathsfbi{\overline{C}}$ on $\Pos_3$.

The geometric structure on $\Pos_3$ supplies us with the natural scalar product to be used in the analysis of linear perturbations.
This scalar product, which depends on $\mathsfbi{\overline{C}}$, is induced by the global distance metric on $\Pos_3$ and is given in (\ref{localMetric}).
If we use the form of the additive perturbation given in (\ref{GTrunc1}), the natural scalar product reduces to the standard Frobenius norm.
By taking the base-point into account, (\ref{localMetric}) also allows us to compare the norm of tangent vectors at \emph{different} base-points.

\section{Evolution of perturbative deformations} 
\label{sec:perturbEvolve}

In this section we derive the evolution equations for the first two fluctuating terms in the weakly nonlinear expansion/deformation of the velocity field/conformation tensor (coefficients of order $\varepsilon$ and $\varepsilon^2$).
Here we assume the governing equations for the dimensionless velocity, $\bld{u}$,  and conformation tensor, $\mathsfbi{C}$, are given by
\begin{align}
	\bld{\nabla} \bcdot \bld{u}  &= 0  \label{govEqn0}  \\
	\dd{\bld{u}}{t} &= -\bnabla p + \frac{\beta}{\Rey}\bld{\Delta} \bld{u} + \frac{1-\beta}{\Rey} \bnabla \bcdot  \mathsfbi{T}   \label{govEqn1}  \\
	\stackrel{\bnabla}{\mathsfbi{C}}  &=  - \mathsfbi{T}  \label{govEqn2} 
\end{align}     
where $\dd{(\cdot)}{t} = \p_t(\cdot) + u_k \p_k(\cdot)$ is the convective derivative,
 $\stackrel{\bld{\nabla}}{\mathsfbi{A}}\equiv \dd{\mathsfbi{A}}{t} -2\,\sym(\mathsfbi{A} \bcdot  \bld{\nabla} \bld{u})$ is the upper-convected Maxwell derivative of $\mathsfbi{A}$, $\sym(\mathsfbi{A}) = \frac{1}{2}(\mathsfbi{A} + \mathsfbi{A}^{\mathsf{T}})$ is the symmetric part of $\mathsfbi{A}$, $p$ is the dimensionless pressure, and $\mathsfbi{T}=\mathsfbi{T}(\mathsfbi{C})$ is the polymer stress. 
In (\ref{govEqn1}), $\Rey = \tau_{\textrm{viscous}}/\tau_{\textrm{inertial}}$ is the Reynolds number, where $\tau_{\textrm{viscous}}$ and $\tau_{\textrm{inertial}}$ are the viscous and inertial time-scales, and $\beta = \nu_{\textrm{solvent}}/\nu_{\textrm{total}}$, where $\nu_{\textrm{solvent}}$ and $\nu_{\textrm{total}}$ are the solvent and total viscosities. 
The functional form  of $\mathsfbi{T}$ depends on the choice of constitutive model. The theoretical development will not depend on this choice but we are interested in models of the form
\begin{align}
	\mathsfbi{T}(\mathsfbi{C}) = \frac{1}{\Wie} \left[f(\tr{\mathsfbi{C}}) \mathsfbi{C}-f(3 )\mathsfbi{I}  \right], \label{OLB_FENEP}
\end{align}
where  $f(s) = [1 - (s/L_{\max}^2)]^{-1}$ for the FENE-P model, and $L_{\max}$ is the maximum extensibility. 
By setting $L_{\max} \rightarrow \infty$ we retrieve the Oldroyd-B model, where $f=1$. 
The Weissenberg number is given by  $\Wie = \tau_{\textrm{relaxation}}/\tau_{\textrm{inertial}}$, where $\tau_{\textrm{relaxation}}$ is the polymer relaxation time.   

We wish to examine the evolution of $(\bld{u},\mathsfbi{C})$ about a given steady solution, $(\bld{\overline{u}},\mathsfbi{\overline{C}})$. 
The velocity field and pressure, cf. (\ref{uexpansion}), are expressed as weakly nonlinear expansions
\begin{align}
	\bld{u} &=  \bld{\overline{u}} + \sum_{k=1}^{N}\varepsilon^{k} \bld{u}_{(k)}. \label{Rey_wnl} \\
	p &= \overline{p} + \sum_{k=1}^{N}\varepsilon^{k} p_{(k)}.  \label{p_wnl}
\end{align} 
and the weakly nonlinear deformation of the conformation tensor is the expansion given in \eqref{nl_expansion1} 
\begin{align}
	\mathsfbi{G}= \mathsfbi{I} + \varepsilon  \bld{\mathcal{G}}_{(1)} +  \varepsilon^2 \left( \frac{\bld{\mathcal{G}}_{(1)}^2}{2} +   \bld{\mathcal{G}}_{(2)}\right) + \hdots \tag{\ref{nl_expansion1}}
\end{align}
where we note, by \eqref{NecessaryDecomposition}, we can re-write \eqref{nl_expansion1} as
\begin{align}
	\mathsfbi{C}= \mathsfbi{\overline{C}} + \varepsilon  \mathsfbi{\overline{F}}\bcdot \bld{\mathcal{G}}_{(1)}\bcdot \mathsfbi{\overline{F}}^{\mathsf{T}} +  \varepsilon^2 \mathsfbi{\overline{F}}\bcdot \left( \frac{\bld{\mathcal{G}}_{(1)}^2}{2} +   \bld{\mathcal{G}}_{(2)}\right)\bcdot \mathsfbi{\overline{F}}^{\mathsf{T}} + \hdots. \label{C_wnd_expansion}
\end{align}

Substituting \eqref{Rey_wnl}, \eqref{p_wnl}, the Taylor series expansion of \eqref{OLB_FENEP}, \eqref{C_wnd_expansion}, all into \eqref{govEqn1}, and equating coefficients of $\varepsilon$ yields the evolution equation for $\bld{u}_{(1)}$,
\begin{multline}
	\p_t \bld{u}_{(1)} + \bld{\overline{u}} \bcdot \bnabla    \bld{u}_{(1)}    +
	\bld{u}_{(1)} \bcdot \bnabla  \bld{\overline{u}} 
	= -\bnabla p_{(1)}  + \frac{\beta}{\Rey}\bld{\Delta} \bld{u}_{(1)} + \frac{1-\beta}{\Wie \Rey} \bnabla \bcdot
	\left[ f(\tr \mathsfbi{\overline{C}})    \mathsfbi{\overline{F}}\bcdot \bld{\mathcal{G}}_{(1)}\bcdot \mathsfbi{\overline{F}}^{\mathsf{T}}  \right]  \\ + \frac{1-\beta}{\Wie \Rey} \bnabla \bcdot
	\left[ \frac{f^2(\tr \mathsfbi{\overline{C}}  )}{L_{\max}^{2}} \tr \left(  \mathsfbi{\overline{F}}^{\mathsf{T}} \bcdot \mathsfbi{\overline{F}}\bcdot \bld{\mathcal{G}}_{(1)} \right)  \mathsfbi{\overline{C}}   \right]. \label{u1_expansion}
\end{multline} 
Similarly, equating coefficients of $\varepsilon^2$ yields the evolution equation for $\bld{u}_{(2)}$,
\begin{multline}
	\p_t \bld{u}_{(2)}  + \bld{\overline{u}} \bcdot \bnabla     \bld{u}_{(2)}  +
	\bld{u}_{(2)} \bcdot \bnabla  \bld{\overline{u}}  +
	\bld{u}_{(1)} \bcdot \bnabla  \bld{u}_{(1)}  
	= -\bnabla p_{(2)}  + \frac{\beta}{\Rey}\bld{\Delta} \bld{u}_{(2)} \\
	+ \frac{1-\beta}{\Wie \Rey} \bnabla \bcdot
	\Bigg[  \mathsfbi{\overline{F}}\bcdot \left( \frac{\bld{\mathcal{G}}_{(1)}^2}{2} +   \bld{\mathcal{G}}_{(2)}\right)\bcdot \mathsfbi{\overline{F}}^{\mathsf{T}}  f(\tr \mathsfbi{\overline{C}})  \Bigg] \\
	+ \frac{1-\beta}{\Wie \Rey} \bnabla \bcdot
	\Bigg\{\frac{f^2(\tr \mathsfbi{\overline{C}})}{L_{\max}^{2}}  \Bigg[   \tr \left( \mathsfbi{\overline{F}}^{\mathsf{T}} \bcdot \mathsfbi{\overline{F}}\bcdot \left( \frac{\bld{\mathcal{G}}_{(1)}^2}{2} +   \bld{\mathcal{G}}_{(2)}\right) \right) \mathsfbi{\overline{C}}    \\
	+  \frac{f(\tr \mathsfbi{\overline{C}})}{L_{\max}^{2}} 
	\tr^2 \left(  \mathsfbi{\overline{F}}^{\mathsf{T}} \bcdot \mathsfbi{\overline{F}}\bcdot \bld{\mathcal{G}}_{(1)} \right) \mathsfbi{\overline{C}}         
	+   \tr \left(  \mathsfbi{\overline{F}}^{\mathsf{T}} \bcdot \mathsfbi{\overline{F}}\bcdot \bld{\mathcal{G}}_{(1)} \right)       \mathsfbi{\overline{F}}\bcdot \bld{\mathcal{G}}_{(1)}\bcdot \mathsfbi{\overline{F}}^{\mathsf{T}} \Bigg]   \Bigg\}. \label{u2_expansion}
\end{multline}

We next consider the expansion of the conformation-tensor equation \eqref{govEqn2}. 
Rather than proceed directly with $\mathsfbi{C}$, we start from the equation for the fluctuating conformation tensor $\mathsfbi{G}$ provided by \cite{Hameduddin2018a}. For a time--invariant base-state, we have
\begin{align}
	\p_t \mathsfbi{G} + \bld{u}\bcdot \bnabla \mathsfbi{G}
	=
	2\, \sym(\mathsfbi{G} \bcdot \mathscr{F}(\bld{u}) )
	- \mathsfbi{M}
	\label{GEvolEqn}
\end{align} 
where  
\begin{align} 
	\mathsfbi{M} &\equiv
	\mathsfbi{\overline{F}}^{-1} \bcdot\mathsfbi{T} \bcdot \mathsfbi{\overline{F}}^{-\mathsf{T}} 
	\label{GStress}  
\end{align}
and we defined the following tensor valued function  
\begin{align}
\mathscr{F}(\bld{a}) = \mathsfbi{\overline{F}}^{\mathsf{T}}\bcdot  \bnabla\bld{a} \bcdot \mathsfbi{\overline{F}}^{-\mathsf{T}}
- 
\left(\mathsfbi{\overline{F}}^{-1}\bcdot\left( \bld{a} \bcdot \bnabla  \right) \mathsfbi{\overline{F}} \right)^{\mathsf{T}}. \label{opFdefn}
\end{align}

  Again substituting \eqref{Rey_wnl}, \eqref{p_wnl},  the Taylor series expansion of \eqref{OLB_FENEP}, \eqref{C_wnd_expansion}, all into \eqref{GEvolEqn}--\eqref{opFdefn}, and equating coefficients of $\varepsilon$ yields an equation for the evolution of $\bld{\mathcal{G}}_{(1)}$, 
\begin{multline}
	\p_t  \bld{\mathcal{G}}_{(1)}  + \bld{\overline{u}}  \bcdot \bnabla \bld{\mathcal{G}}_{(1)}  
	=
	2 \, \sym \left[ \mathscr{F}(\bld{u}_{(1)})      
	+   \bld{\mathcal{G}}_{(1)}  \bcdot  \mathscr{F}(\bld{\overline{u}})  \right]  - \frac{1}{\Wie} f(\tr \mathsfbi{\overline{C}}) \bld{\mathcal{G}}_{(1)}  \\
	-  \frac{f^2(\tr \mathsfbi{\overline{C}})}{L_{\max}^{2}\Wie}  \tr \left(  \mathsfbi{\overline{F}}^{\mathsf{T}} \bcdot \mathsfbi{\overline{F}}\bcdot \bld{\mathcal{G}}_{(1)} \right)   \mathsfbi{I}.   \label{G1_expansion}
\end{multline}    
Similarly,  \eqref{Rey_wnl}, \eqref{p_wnl},  the Taylor series expansion of \eqref{OLB_FENEP}, \eqref{C_wnd_expansion}, all into \eqref{GEvolEqn}--\eqref{G1_expansion}, and equating coefficients of $\varepsilon^2$ yields an equation for the evolution of $\bld{\mathcal{G}}_{(2)}$ as
\begin{multline}
	\p_t  \bld{\mathcal{G}}_{(2)}+   \bld{\overline{u}}  \bcdot \bnabla\bld{\mathcal{G}}_{(2)}
	= 
	2\,  \sym [ \mathscr{F}(\bld{u}_{(2)} )+ \bld{\mathcal{G}}_{(2)}\bcdot \mathscr{F}(\bld{\overline{u}})  ]    - \frac{f(\tr \mathsfbi{\overline{C}})}{ \Wie}  \bld{\mathcal{G}}_{(2)}\\ - \bld{u}_{(1)}   \bcdot \bnabla     \bld{\mathcal{G}}_{(1)}  -\bld{\mathcal{G}}_{(1)} \bcdot \sym [\mathscr{F}(\bld{\overline{u}})] \bcdot  \bld{\mathcal{G}}_{(1)}  \\ +
	\bld{\mathcal{G}}_{(1)} \bcdot  \asym  [\mathscr{F}(\bld{u}_{(1)})]  
	- \asym [ \mathscr{F}(\bld{u}_{(1)}) ]  \bcdot \bld{\mathcal{G}}_{(1)}        
	+\frac{f(\tr \mathsfbi{\overline{C}})}{2 \Wie}  \bld{\mathcal{G}}_{(1)}^2   \\ 
	- \frac{f^2(\tr \mathsfbi{\overline{C}})}{L_{\max}^{2}\Wie}  \Bigg[    
	\tr \left( \mathsfbi{\overline{F}}^{\mathsf{T}} \bcdot \mathsfbi{\overline{F}}\bcdot \left( \frac{\bld{\mathcal{G}}_{(1)}^2}{2} +   \bld{\mathcal{G}}_{(2)}\right) \right)  
	+  \frac{f(\tr \mathsfbi{\overline{C}})}{L_{\max}^{2}} 
	\tr^2 \left(  \mathsfbi{\overline{F}}^{\mathsf{T}} \bcdot \mathsfbi{\overline{F}}\bcdot \bld{\mathcal{G}}_{(1)} \right)     \Bigg]      \mathsfbi{I} \label{G2_expansion}
\end{multline} 
where $\asym(\mathsfbi{A}) = \frac{1}{2}(\mathsfbi{A} - \mathsfbi{A}^{\mathsf{T}})$ is the asymmetric part of the tensor $\mathsfbi{A}$.

By setting $L_{\max} \rightarrow \infty$, we can retrieve the relevant Oldroyd-B equations. 
Thus, for the Oldroyd-B model, the equation for  $\bld{u}_{(1)}$ in \eqref{u1_expansion} reduces to
\begin{multline}
	\p_t \bld{u}_{(1)} + \bld{\overline{u}} \bcdot \bnabla    \bld{u}_{(1)}    +
	\bld{u}_{(1)} \bcdot \bnabla  \bld{\overline{u}} 
	= -\bnabla p_{(1)}  + \frac{\beta}{\Rey}\bld{\Delta} \bld{u}_{(1)} \\ + \frac{1-\beta}{\Wie \Rey} \bnabla \bcdot
	\left[ f(\tr \mathsfbi{\overline{C}})    \mathsfbi{\overline{F}}\bcdot \bld{\mathcal{G}}_{(1)}\bcdot \mathsfbi{\overline{F}}^{\mathsf{T}}  \right], \label{u1_expansion_OLB}
\end{multline} 
the equation for  $\bld{u}_{(2)}$ in \eqref{u2_expansion}  reduces to
\begin{multline}
	\p_t \bld{u}_{(2)}  + \bld{\overline{u}} \bcdot \bnabla     \bld{u}_{(2)}  +
	\bld{u}_{(2)} \bcdot \bnabla  \bld{\overline{u}}  +
	\bld{u}_{(1)} \bcdot \bnabla  \bld{u}_{(1)}  
	= -\bnabla p_{(2)}  + \frac{\beta}{\Rey}\bld{\Delta} \bld{u}_{(2)} \\
	+ \frac{1-\beta}{\Wie \Rey} \bnabla \bcdot
	\Bigg[  \mathsfbi{\overline{F}}\bcdot \left( \frac{\bld{\mathcal{G}}_{(1)}^2}{2} +   \bld{\mathcal{G}}_{(2)}\right)\bcdot \mathsfbi{\overline{F}}^{\mathsf{T}}  f(\tr \mathsfbi{\overline{C}})  \Bigg],    \label{u2_expansion_OLB}
\end{multline} 
the equation for  $ \bld{\mathcal{G}}_{(1)}$ in \eqref{G1_expansion}  reduces to
\begin{align}
	\p_t  \bld{\mathcal{G}}_{(1)}  + \bld{\overline{u}}  \bcdot \bnabla \bld{\mathcal{G}}_{(1)}  
	=
	2 \, \sym \left[ \mathscr{F}(\bld{u}_{(1)})      
	+   \bld{\mathcal{G}}_{(1)}  \bcdot  \mathscr{F}(\bld{\overline{u}})  \right]  
	- \frac{1}{\Wie}   \bld{\mathcal{G}}_{(1)},    \label{G1_expansion_OLB}
\end{align} 
and finally, the equation for  $ \bld{\mathcal{G}}_{(2)}$ in \eqref{G2_expansion}  reduces to
\begin{multline}
	\p_t  \bld{\mathcal{G}}_{(2)}+   \bld{\overline{u}}  \bcdot \bnabla\bld{\mathcal{G}}_{(2)}
	= 2\,  \sym [ \mathscr{F}(\bld{u}_{(2)} )+ \bld{\mathcal{G}}_{(2)}\bcdot \mathscr{F}(\bld{\overline{u}})  ] 
	- \frac{1}{ \Wie}   \bld{\mathcal{G}}_{(2)}  \\- \bld{u}_{(1)}   \bcdot \bnabla     \bld{\mathcal{G}}_{(1)}  -\bld{\mathcal{G}}_{(1)} \bcdot \sym [\mathscr{F}(\bld{\overline{u}})] \bcdot  \bld{\mathcal{G}}_{(1)}  \\ +
	\bld{\mathcal{G}}_{(1)} \bcdot  \asym  [\mathscr{F}(\bld{u}_{(1)})]  
	- \asym [ \mathscr{F}(\bld{u}_{(1)}) ]  \bcdot \bld{\mathcal{G}}_{(1)}      
	+ \frac{1}{2 \Wie}    \bld{\mathcal{G}}_{(1)}^2. \label{G2_expansion_OLB}
\end{multline}

The following skew-symmetric term appears in the equations for $\bld{\mathcal{G}}_{(2)}$, \eqref{G2_expansion} and \eqref{G2_expansion_OLB},
\begin{align}
	\bld{\mathcal{G}}_{(1)} \bcdot  \asym  [\mathscr{F}(\bld{u}_{(1)})]  
	- \asym [ \mathscr{F}(\bld{u}_{(1)}) ]  \bcdot \bld{\mathcal{G}}_{(1)}. \label{volpres_G2}
\end{align}
Since the term is skew-symmetric, it has zero trace and therefore does not contribute towards the evolution of $\tr \bld{\mathcal{G}}_{(2)}$ and thus represents a volume-preserving deformation.
The latter fact can be shown by taking the logarithm of the weakly nonlinear deformation \eqref{nl_expansion0} and, without loss of generality, assuming $N=2$ to obtain
\begin{align}
	\log \det \mathsfbi{G} &=  
	\det(\e{ \varepsilon \bld{\mathcal{G}}_{(1)}/2} \bcdot 
	\e{ \varepsilon^{2} \bld{\mathcal{G}}_{(2)}}  
	\bcdot \e{ \varepsilon \bld{\mathcal{G}}_{(1)}/2}) \\
	&= 
	\log \det \e{ \varepsilon \bld{\mathcal{G}}_{(1)}/2} + 
	\log \det \e{ \varepsilon^{2} \bld{\mathcal{G}}_{(2)}}  +
	\log \det   \e{ \varepsilon \bld{\mathcal{G}}_{(1)}/2}   \\ 
	&=    \varepsilon \tr \bld{\mathcal{G}}_{(1)}  + 
	\varepsilon^{2} \tr \bld{\mathcal{G}}_{(2)}.    \label{logdetG_wnd}
\end{align}  
If $\tr \bld{\mathcal{G}}_{(2)}=0$, the deformation associated with $\bld{\mathcal{G}}_{(2)}$ does not contribute to $\log \det \mathsfbi{G} $ and is volume-preserving.

We will be particularly interested in the case of linear perturbations. Here we set
\begin{align}
	\bld{\mathcal{G}}_{(k)}  =  \bld{0}, \qquad k > 1
\end{align}
and thus for this special case, we have
\begin{align}
	\bld{u} 	&= \bld{\overline{u}} + \bld{u}' 	\approx  \bld{\overline{u}} +  \varepsilon  \bld{u}_{(1)}.   \\
	p 		&= \overline{p} + p' 			\approx  \overline{p} +  \varepsilon  p_{(1)}   \\
	\mathsfbi{G}   &= \e{\bld{\mathcal{G}}} 	\approx \mathsfbi{I} +  \varepsilon \bld{\mathcal{G}}_{(1)}  
\end{align}
The state variables in the linearized equations, (\ref{u1_expansion}) and (\ref{G1_expansion}), are then a velocity field and a tangent to $\Pos_3$, unlike in the full governing equations, \eqref{govEqn0}--\eqref{govEqn2}, where the state variables are a velocity field and conformation tensor field. This is important to note, since tangents to $\Pos_3$ have a distinct interpretation from $\mathsfbi{C}$ and are in $\Sym_3$; they are not required to be positive--definite.

The tangent to $\Pos_3$ in the equations has been expressed using $\bld{\mathcal{G}}$ but, by (\ref{GTrunc1}), it can equivalently expressed using $\mathsfbi{C}'$.
Such linearized equations in terms of $\mathsfbi{C}'$  have been derived previously by directly applying an additive decomposition to the governing equations \citep{Zhang2013,Lee2017}.
The present work expresses the perturbation equations in terms of $\bld{\mathcal{G}}$ because then the scalar product on the local tangent space on $\Pos_3$ coincides with the standard Frobenius scalar product.
Such a formulation is important when the scalar product is needed.
As an example, consider the eigenmodes associated with linearized equations, (\ref{u1_expansion}) and (\ref{G1_expansion}).
These modes are equivalent in both approaches, because the eigenvalue problem (see e.g. Zhang \etal \cite{Zhang2013}) does not depend on the scalar product.
However, projection of a flow state on one of the modes depends on the scalar product.
It follows that projections using the function space generalization of the Frobenius scalar product are most appropriate when we use $\bld{\mathcal{G}}$, and not $\mathsfbi{C}'$, since then the scalar product is consistent with the global metric on $\Pos_3$.
We will be considering the evolution of such modes in the present work. 
We first derive a simple kinematic constraint on the linear evolution of the modes, which arises due to the constraint of positive--definiteness on $\mathsfbi{G}$.

\subsection{Constraint on linear evolution}
\label{sec:linEvolCon}

An initial condition consisting of a small-amplitude unstable mode will initially amplify exponentially as predicted by linear theory. Eventually, however, nonlinear effects will become significant because otherwise the conformation tensor will lose positive--definiteness.
It is of interest to determine an estimate of the maximum time that the evolution of the perturbation can be well approximated by linear theory, i.e.\,along Euclidean lines.
Such an estimate is a useful guide for selecting initial perturbation amplitude, e.g.\,by ruling out initial perturbations that have unacceptably small time for which the linear evolution holds.

Consider an initial condition which is a perturbed base flow, $\bld{u}|_{t=0} = \bld{\overline{u}} + \varepsilon \bld{q}$, $\mathsfbi{C}|_{t=0}= \overline{\mathsfbi{C}} + \varepsilon   \mathsfbi{Q} $ with $\varepsilon \ll 1$, and where $(\bld{q},\mathsfbi{Q})$ is an unstable  eigenmode of the linear stability equations, with growth rate $\omega_{\textrm{i}} > 0$.
For the purpose of the current derivation, it is helpful to re-write the conformation tensor by pre-multiplying $\mathsfbi{C}|_{t=0}$ by $\mathsfbi{\overline{F}}^{-1}$ and post-multiplying by $\mathsfbi{\overline{F}}^{-\mathsf{T}}$ (see equation \ref{NecessaryDecomposition}).  
This operation yields $\mathsfbi{G}|_{t=0} =  \mathsfbi{I} + \varepsilon \bld{\mathcal{Q}}$, where $\bld{\mathcal{Q}} = \mathsfbi{\overline{F}}^{-1} \bcdot \mathsfbi{Q} \bcdot  \mathsfbi{\overline{F}}^{-\mathsf{T}}$. 
If we assume that the mode grows according to linear theory for some time and $\mathsfbi{G}$ evolves along Euclidean lines, then $\mathsfbi{G}(t) = \mathsfbi{I} + \varepsilon \bld{\mathcal{Q}} \e{\omega_{\textrm{i}} t}$. 
If any eigenvalue of $\bld{\mathcal{Q}}$ is negative, $\mathsfbi{I} + \varepsilon \bld{\mathcal{Q}} \e{\omega_{\textrm{i}} t}$ will eventually lose positive--definiteness.

Suppose $\bld{\mathcal{Q}}$ is not zero and is harmonic in a spatial direction, then $\bld{\mathcal{Q}}$ has a strictly negative eigenvalue somewhere in the domain.
For positive--definiteness of $\mathsfbi{G}$, we require $1 + \varepsilon  \sigma_i\,(\bld{\mathcal{Q}}) \e{\omega_{\textrm{i}} t} > 0$ for each $i=1,2,3$, where  $\sigma_i\,(\bld{\mathcal{Q}}) $ denotes the $i$-th largest eigenvalue of the tensor $\bld{\mathcal{Q}}$.
Wherever $\sigma_i\,(\bld{\mathcal{Q}}) < 0$, the dynamics must  induce a curvature on the evolution along $\Pos_3$ before a time $t_{\max}$ when the eigenvalue crosses zero. This $t_{\max}$ is given by
\begin{align}
	\omega_{\textrm{i}} t_{\max}   =   -\left(\log\,\varepsilon  + \log  \max_i |\sigma_i (\bld{\mathcal{Q}})|\right), \label{TPosCon}
\end{align}
and determines an upper bound on the time for which evolution of $\mathsfbi{G}$ along Euclidean lines does not violate the positive definiteness constraint on the conformation tensor.
The condition (\ref{TPosCon}) is a guide for selecting the initial perturbation amplitude $\varepsilon$, based on $t_{\max}$;
reducing $\varepsilon$, one can arbitrarily increase $t_{\max}$ to the desired value.

Instead of evolving along Euclidean lines, if one assumes that the perturbation evolves along a geodesic, then $\mathsfbi{G} = \e{\varepsilon\bld{\mathcal{G}}_{(1)}} $ as in equation (\ref{ItoGPath}).
Such an evolution remains on the manifold of $\Pos_3$ for any perturbation amplitude.  
We would then formally have the superexponential evolution $\mathsfbi{G} = \e{\varepsilon\bld{\mathcal{Q}}\e{\omega_{\textrm{i}} t}}$. 
Expanding the exponentials it can be easily shown that such an evolution is equivalent to evolution along Euclidean lines for sufficiently small $\varepsilon$. 

Physically, a perturbation that is harmonic in space leads to regions of the flow where the polymers are compressed much more, in the sense of a volumetric change, than the maximum expansion. This is because positive and negative additive perturbations to $\mathsfbi{\overline{C}}$  with equal magnitudes are not of equal magnitude with respect to the natural distance on $\Pos_3$.


\section{Tollmien--Schlichting waves in viscoelastic channel flow }  
 \label{sec:DNS}
 
\begin{figure}
	\centering
	\includegraphics[width=0.55\textwidth]{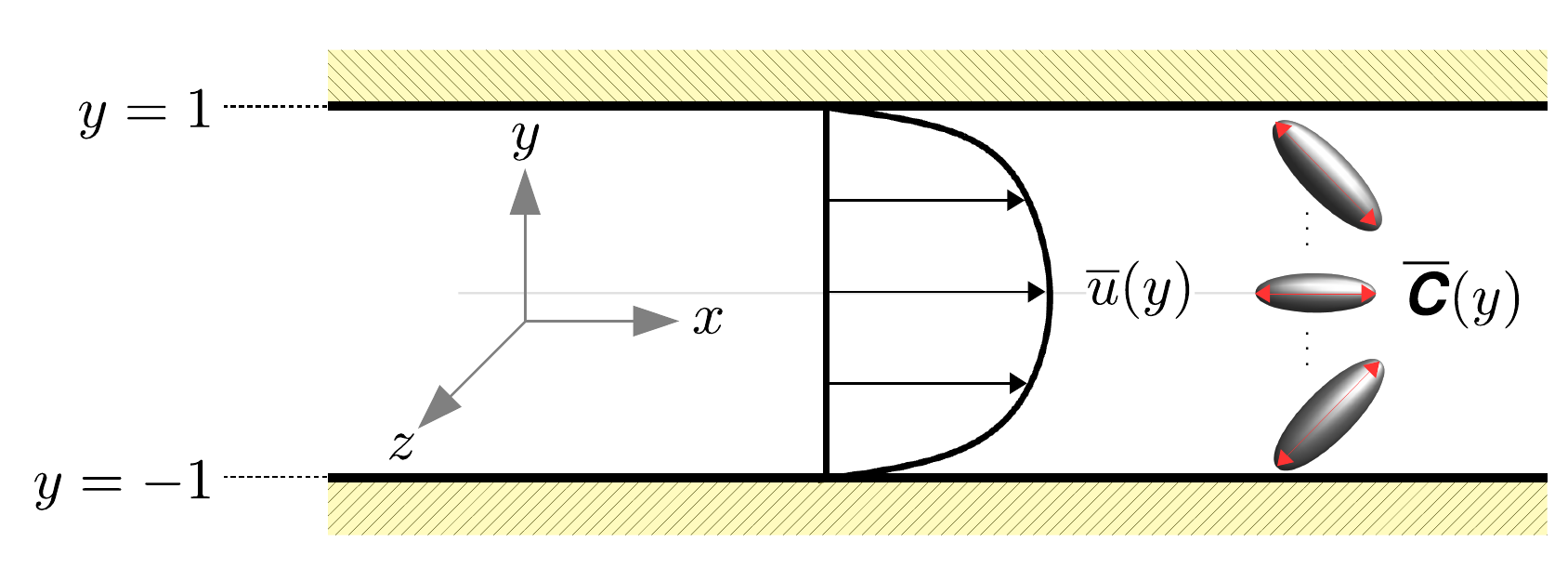}
	\caption{Schematic of viscoelastic channel flow and the laminar base flow. The base flow is uniform in $x$ and $z$. The Tollmien-Schlichting waves are uniform in the $z$ direction.}
\label{fig:channel_flow_config}
\end{figure}

In this section, we use direct numerical simulations (DNS) to examine the nonlinear evolution of Tollmien--Schlichting (TS) wave in channel flow of a FENE-P fluid.
The stress relation for a FENE-P fluid was provided in (\ref{OLB_FENEP}).
The geometry of the channel flow setup, along with the laminar base flow, are sketched in figure \ref{fig:channel_flow_config}.
The base flow is only a function of the wall-normal coordinate ($y$), and the 
Tollmien\rev{--}Schlichting waves are independent of the spanwise coordinate ($z$).
In the present viscoelastic case, the waves are unstable eigenmodes of the linearized equations (\ref{u1_expansion}) and (\ref{G1_expansion}).

Results from the DNS will be used to illustrate some of the theoretical developments described in the previous sections. 
In \S\ref{sec:TS:bg}, we provide the background and motivation for studying this particular problem. 
The simulation setup and the laminar base flow are described in \S\ref{sec:TS:simulation}.   
In \S\ref{sec:TS:initial} we discuss the initial perturbation from the geometric viewpoint, and the predicted upper bound for the duration of purely linear evolution along Euclidean lines (see also \S\ref{sec:linEvolCon}).
The development of the Tollmien--Schlichting wave is considered in detail in \S\ref{sec:TS:time} and, finally, we utilize the weakly nonlinear deformation framework (see \S\ref{sec:wnd}) to study how the Tollmien-Schlichting waves first deviate from exponential growth.

\subsection{Background}

\label{sec:TS:bg}

The growth and saturation of Tollmien--Schlichting waves in channel flow considered here was simulated previously by \citet{Lee2017}.
That study motivates the present focus on TS waves and, for this reason, its main relevant findings are outlined.

\citet{Lee2017} performed linear stability analyses of channel flow of FENE-P fluid, with $\beta=0.90$ and $L_{\max}=100$.  
They reported the growth rates of the instability waves in $(\Rey,\Wie)$-space, and identified the neutral curve which exhibited non-monotonic behaviour:  
The critical $\Rey$ initially decreased and subsequently increased with increasing $\Wie$. 
At $\Rey=4667$, they examined the two-dimensional mode which is most unstable in the Newtonian configuration, and showed that its growth rate increased with Weissenberg number up to a maximum at $\Wie \approx 1.83$.
Further increasing $\Wie$ to $4.50$ decreased the maximum growth rate to the Newtonian value, and even lower for $\Wie>4.50$.  
The authors simulated the nonlinear evolution of Tollmien--Schlichting waves at $\Rey=4667$ for Newtonian fluid and viscoelastic fluids at $\Wie\in\{1.83,4.50,6.67\}$, i.e.\,cases with modal growth rates below, at, and above the Newtonian value.

\cite{Lee2017} examined the evolution of the modes using the (spatially averaged) perturbation kinetic energy and found, for all cases, that it initially grew at a rate consistent with predictions from linear theory.
In all but the cases with $\Wie=4.50$, the kinetic energy eventually saturated; our focus is on conditions where an equilibrium solution is possible, and hence $\Wie=4.50$ is not discussed further.
The $\Wie=1.83$ and $\Wie=6.67$ will hereafter be referred to as $W1.83$ and $W6.67$.
The saturation energy for the Newtonian case and $W1.83$ coincided, while $W6.67$ saturated at a much lower energy.
Superficially, the saturation appeared to be similar, but \citet{Lee2017} found that the behaviour of the flow before and after saturation was distinctly different in $W1.83$ compared to $W6.67$.

Just prior to saturation, \citet{Lee2017} reported that the growth rate of the average kinetic energy in $W1.83$ increases substantially beyond the linear rate\textemdash qualitatively similar to behaviour in the Newtonian case.
Qualitative similarity was also reported for the mean-velocity distortion as well as the term-by-term breakdown of the perturbation kinetic-energy budget in the saturated state.
The flow behaviour for $W6.67$ was markedly different: 
there was no increase in the growth rate of average kinetic energy prior to saturation; 
the mean-velocity distortion in the saturated state was astonishingly small; 
the term-by-term breakdown of the perturbation kinetic-energy budget in the saturated state bore more resemblance to the linear growth stage, such as predominance of the critical-layer peak in the production term, rather than to the saturated states for $W1.83$ and the Newtonian case.
We refer the reader to the original paper \citep{Lee2017} for detailed figures documenting these distinctions.
In the present work, we will leverage our new framework to shed more light on the differences between the polymer deformation in the two cases $W1.83$ and $W6.67$ as they saturate, as opposed to focusing on the velocity field and its modification by the polymer, which was the focus of \citet{Lee2017}.

\subsection{Simulation setup}
\label{sec:TS:simulation}

The setup of the direct numerical simulations performed herein is identical to the one described by \citet{Lee2017}.  
The flow in the channel is maintained at a constant mass rate with bulk Reynolds number $\Rey\equiv h U_b / \nu =4667$, where $U_b$ is the bulk velocity, $h$ is the channel half height and $\nu$ is the total kinematic viscosity of the fluid.  
The two conditions simulated correspond to $\Wie \equiv \tau_{\textrm{relaxation}} U_b / h \in\{1.83, 6.67\}$, both with $\beta = 0.9$ and $L_{\max}=100$.
The flow is composed of a laminar base state at these parameters, plus a small-amplitude two-dimensional Tollmien-Schlichting wave.

The domain and grid sizes are listed in table \ref{tab:params}. 
The simulations assume all fields are periodic in the streamwise $x$ and spanwise $z$ directions, and imposes no-slip condition on the velocity field at the two walls ($y=\pm 1$).
There are no boundary conditions on the conformation tensor.
The numerical scheme used in the present simulations is similar to that used by \citet{Lee2017}, except that here we use second-order Runge--Kutta time-stepping for the conformation tensor and a variant of the slope-limiting approach developed by \citet{Vaithianathan2006} for the conformation tensor advection. 
The latter is designed to ensure positive--definiteness of the conformation tensor (see \S2.2 by \citet{Lee2017} and appendix B by \citet{Hameduddin2018a}).
The algorithm has been extensively validated, by comparing exponential growth rates of instability waves to predictions from linear theory \citep[][and present study]{Lee2017} and non-modal amplification of disturbances \citet{Agarwal2014}.  
 
\begin{table}
	\begin{center}
		\begin{tabular}{cccccccccc} 
			\multicolumn{6}{c}{ } & \multicolumn{2}{c}{Eigenvalue, $\omega_{\textrm{r}}+\textrm{i}\omega_{\textrm{i}}$ } & Domain size & Grid size  \\ 
		Case 	& $\Wie$ & $\beta$ & $L_{\max}$   & $\Rey$ & $k_x$ & $\omega_{\textrm{r}}$ & $\omega_{\textrm{i}}$ & $L_{x}\times L_{y} \times L_{z}$& $N_{x}\times N_{y} \times N_{z}$   \\ 
		$W1.83$ 	& $1.83$ & $0.90$ & $100$  & $4667$ & $1.00$ & $0.3792$ & $3.489 \times 10^{-3}$  & $2\pi \times 2 \times 0.1$ & $160\times 2048 \times 16$	 			\\
		$W6.67$ 	&$6.67$ & $0.90$ & $100$   &	 $4667$ & $1.00$ & $0.3799$ & $1.571 \times 10^{-3}$   & $2\pi \times 2 \times 0.1$ & $160\times 2048 \times 16$	    
		\end{tabular}
	\end{center}
	\caption{Parameters of the simulation setup and the viscoelastic Tollmien--Schlichting wave. The characteristic length and velocity are the channel half-height and bulk flow speed.}
	\label{tab:params}
\end{table}

\begin{figure}
  \centering
  \includegraphics[width=0.5\textwidth]{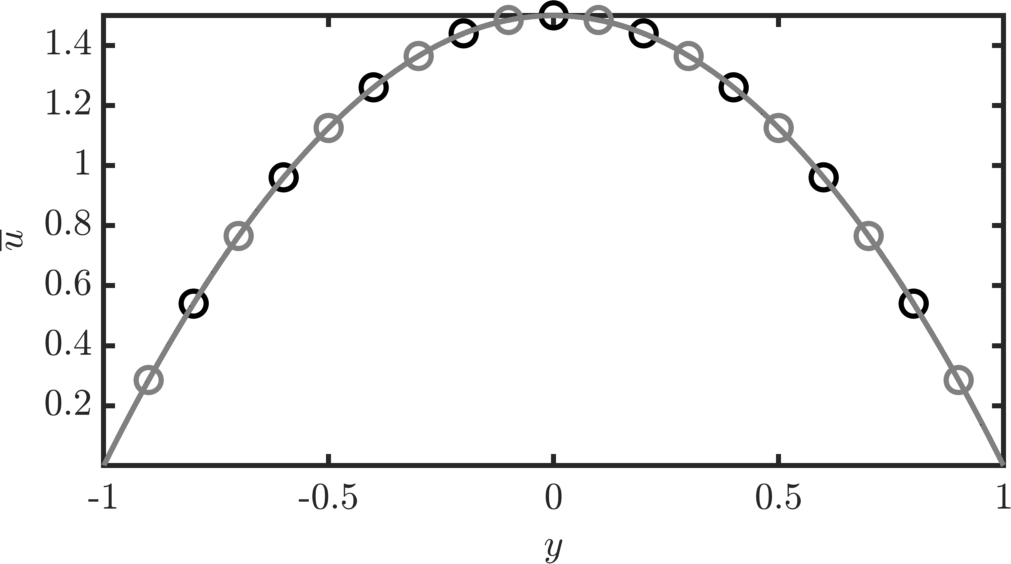}
  \caption{Laminar streamwise velocity profile for $W1.83$ (black line with symbols, \linesolidsSymCirc) and $W6.67$ (grey line with symbols, {\color{gray}\linesolidsSymCirc}).}
  \label{fig:u_laminar}
\end{figure}

The laminar base state can be derived from the governing equations by assuming that the flow is fully developed, so that the state variables are only functions of $y$. 
The $x$, $y$, and $z$ components of the laminar velocity field, respectively, are given by
\begin{align}
&\overline{u}(y) = \frac{1}{2} \frac{\Rey}{\beta}\frac{\text{d}\overline{p}}{\text{d}x} (y^2 -1) - \frac{1-\beta}{\beta} \int_{-1}^{y} \overline{\mathsfi{T}}_{xy}(s)\,\text{d}s, \quad
\overline{v} = 0, \quad
\overline{w} = 0
\end{align}
and the laminar pressure gradient $\text{d}\overline{p}/\text{d}x$ is a fixed constant chosen so that the bulk velocity is unity, $\frac{1}{2} \int_{-1}^{1}\overline{u}(y)\,\text{d}y = 1$.
The laminar velocity profiles for $\Rey=4667$, $\beta=0.90$, $L_{\max}=100$, and $\Wie\in \{1.83,6.67\}$ are shown in figure \ref{fig:u_laminar}.
The two curves are indistinguishable, and are as similar to the Newtonian profile as they are to one another.

\begin{figure}
  \centering
  \includegraphics[width=0.8\textwidth]{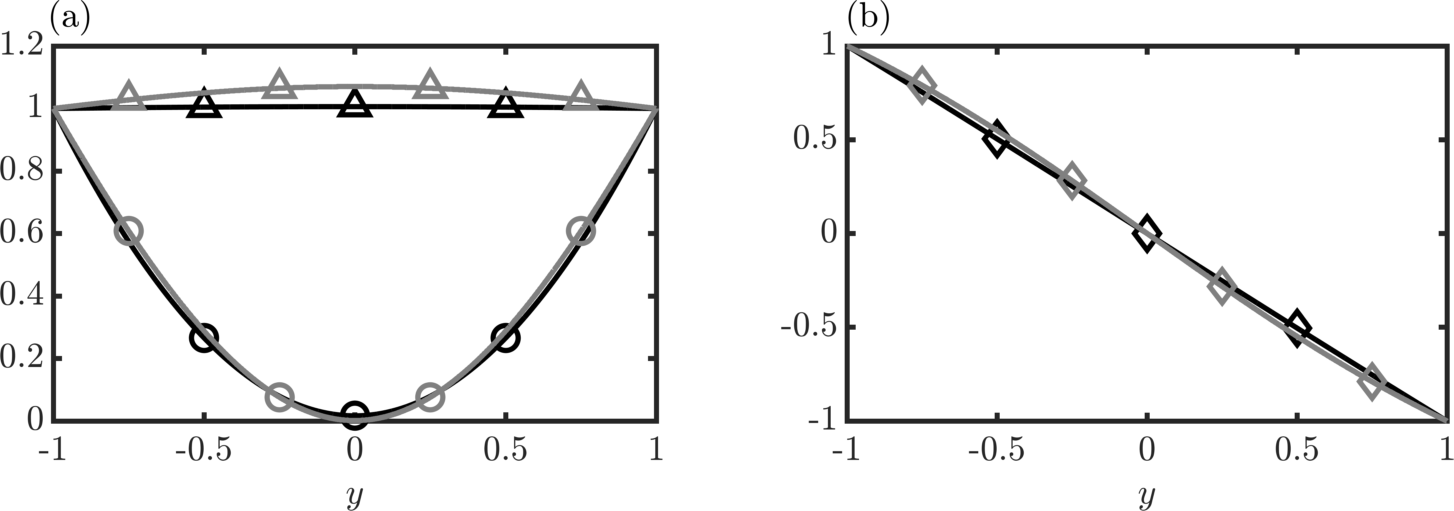}
  \caption{Components of the laminar base flow conformation tensor $\mathsfbi{\overline{C}}$, each normalized by its value at the bottom wall ($y=-1$). Black lines are for $W1.83$ and grey lines are for $W6.67$. (a) $\mathsfi{\overline{C}}_{xx}$ (\linesolidsSymCirc, {\color{gray}  \linesolidsSymCirc }) and $\mathsfi{\overline{C}}_{yy}$ (\linesolidsSymTriangleUp, {\color{gray}  \linesolidsSymTriangleUp }). (b)  $\mathsfi{\overline{C}}_{xy}$ (\linesolidsSymDiamond, {\color{gray}  \linesolidsSymDiamond }). For $W1.83$, $\mathsfi{\overline{C}}_{xx}|_{y=-1}=60.15$, $\mathsfi{\overline{C}}_{yy}|_{y=-1}=0.99$, and $\mathsfi{\overline{C}}_{xy}|_{y=-1}=5.42$.
      For $W6.67$, $\mathsfi{\overline{C}}_{xx}|_{y=-1}=656.48$, $\mathsfi{\overline{C}}_{yy}|_{y=-1}=0.93$, and $\mathsfi{\overline{C}}_{xy}|_{y=-1}=17.50$.
}
  \label{fig:Cij_laminar}
\end{figure}

The polymer stress component $\overline{\mathsfi{T}}_{xy}$ is a solution to the following cubic equation,
\begin{align}
&\frac{1}{L_{\max}^2}\overline{\mathsfi{T}}_{xy}^3 + \frac{f(3)}{2\Wie^2} \left(1 + \frac{2f(3)+1}{L_{\max}^2} + \frac{1-\beta}{\beta}f(3) \right)\overline{\mathsfi{T}}_{xy}
- \frac{\Rey f^2(3)}{2\beta \Wie^2 }  \frac{\text{d}\overline{p}}{\text{d}x}y = 0. \label{lamTxyeqn}
\end{align}
The discriminant of (\ref{lamTxyeqn}) is negative and thus there is only one real root, which can be obtained using standard methods such as the analytical approach by \citet{Cardano1993}.
The remaining components of the stress are
\begin{align}
&\overline{\mathsfi{T}}_{xx} = \frac{2\Wie}{f(3)} \overline{\mathsfi{T}}_{xy}^2, \quad \overline{\mathsfi{T}}_{yy}=\overline{\mathsfi{T}}_{zz} = \overline{\mathsfi{T}}_{xz} = \overline{\mathsfi{T}}_{yz} = 0, 
\end{align}
and the associated conformation tensor can be calculated using the stress relation (\ref{OLB_FENEP}).
Figure \ref{fig:Cij_laminar} shows the non-zero components of the laminar base-state conformation tensor, normalized so their magnitudes are comparable.
The normalized profiles are very similar but their absolute magnitudes are dramatically different, highlighting the importance of correctly ascertaining the relative size of a perturbation to these profiles.


The two-dimensional initial perturbation is evaluated separately by solving the eigenvalue problem associated with the stability of the laminar base flow. 
The streamwise wavenumber is prescribed, $k_x = 1$, and the eigenmode with the highest growth rate is selected at each flow condition.
The modal complex frequencies, $\omega_{\textrm{r}} + \textrm{i}\omega_{\textrm{r}}$, are listed in table \ref{tab:params}. 
The temporal growth rate is given by the complex part of the eigenvalue, $\omega_{\textrm{i}}$, and the phase velocity is given by $\omega_{\textrm{r}}/k_x$.

\subsection{Initial perturbation}

\label{sec:TS:initial}

\begin{figure}
	{	\centering
                    \parbox{\textwidth}{
      \hspace{0.5in} (a)
      \hspace{1in} (b)
      \hspace{1in} (c)
      \hspace{1in} (d)
    }
		
		\includegraphics[scale=1.0,trim={0.in 0.2in 0in 0.2in},clip]{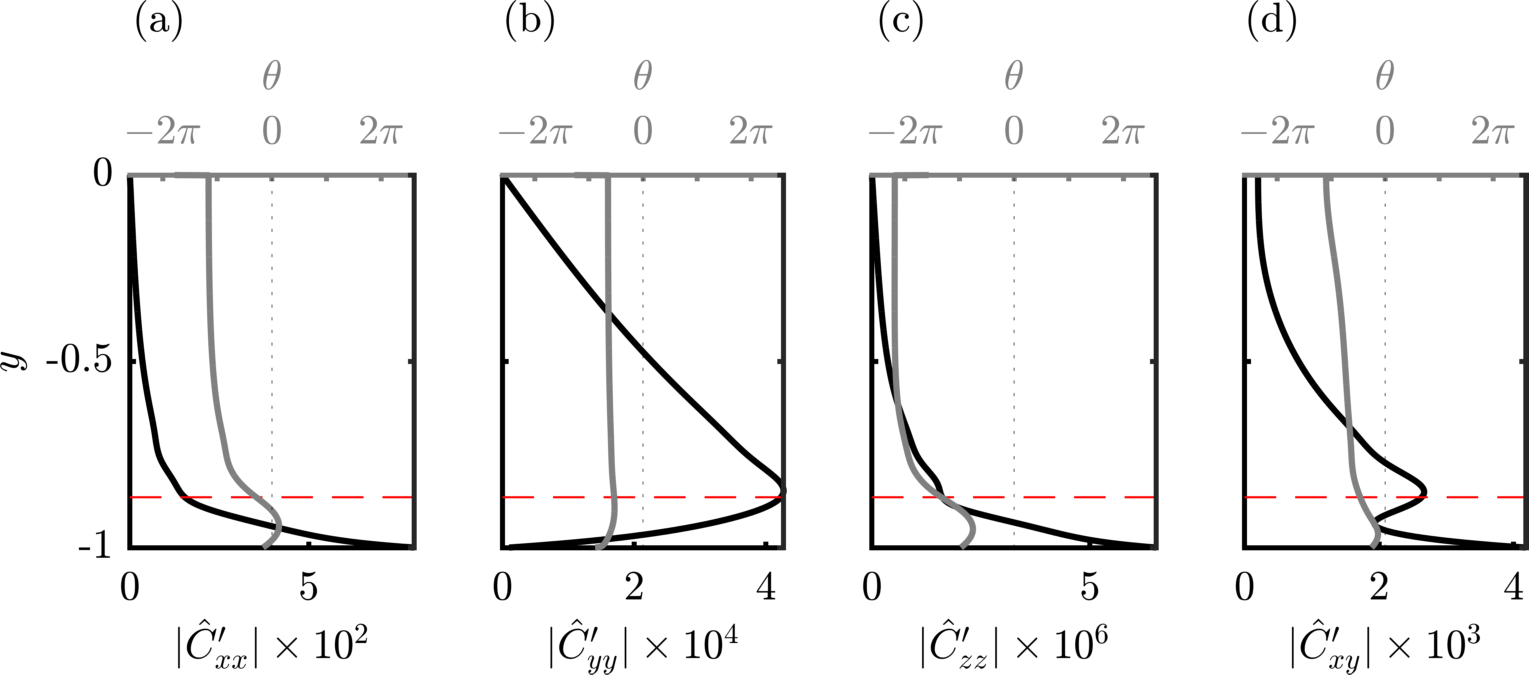}
    
                \parbox{\textwidth}{
      \hspace{0.65in} $|\mathsfi{\hat{C}}_{xx}'|\times 10^2$ 
      \hspace{0.55in} $|\mathsfi{\hat{C}}_{yy}'|\times 10^4$
      \hspace{0.55in} $|\mathsfi{\hat{C}}_{zz}'|\times 10^6$
      \hspace{0.55in} $|\mathsfi{\hat{C}}_{xy}'|\times 10^3$
    }	
		
		\vspace{0.2in}
		
		\centering
                \parbox{\textwidth}{
      \hspace{0.5in} (e)
      \hspace{1in} (f)
      \hspace{1in} (g)
      \hspace{1in} (h)
    }
  
	\includegraphics[scale=1.0,trim={0.in 0.2in 0in 0.2in},clip]{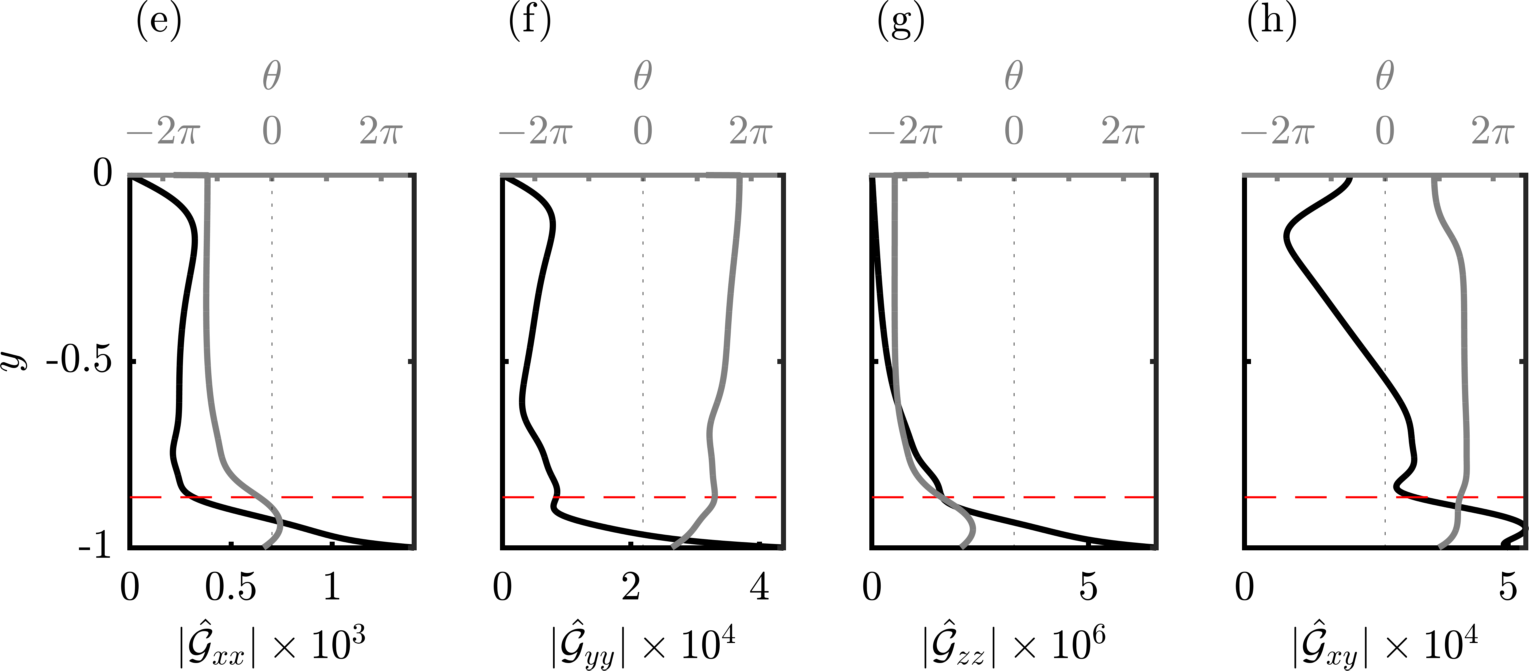}	
    
            \parbox{\textwidth}{
      \hspace{0.65in} $|\mathcal{\hat{G}}_{xx}|\times 10^3$ 
      \hspace{0.55in} $|\mathcal{\hat{G}}_{yy}|\times 10^4$
      \hspace{0.55in} $|\mathcal{\hat{G}}_{zz}|\times 10^6$
      \hspace{0.55in} $|\mathcal{\hat{G}}_{xy}|\times 10^4$
    }	
	}
	
	\caption{$\Wie=1.83$. Components of the initial perturbation tensor Fourier mode:  (a)--(d) in the native form, $\hat{\mathsfbi{C}}'|_{t=0}$, and (e)--(h) in the form of a tangent on $\Pos_3$, $\bld{\hat{\mathcal{G}}}|_{t=0}$. In all panels: solid black lines are the absolute magnitudes of the modes, solid grey lines are the phase angles $\theta$. The horizontal red dashed line is the location of the critical layer, and the vertical thin black dotted line marks zero phase angle. Note: As indicated in the axes labels, the perturbation values are normalized to optimize the clarity of the plots.} 
	
	\label{fig:Ceigenmode_1}
\end{figure}

\begin{figure}
	{	\centering
        \parbox{\textwidth}{
      \hspace{0.5in} (a)
      \hspace{1in} (b)
      \hspace{1in} (c)
      \hspace{1in} (d)
    }
  
    \includegraphics[scale=1.0,trim={0.in 0.2in 0in 0.2in},clip]{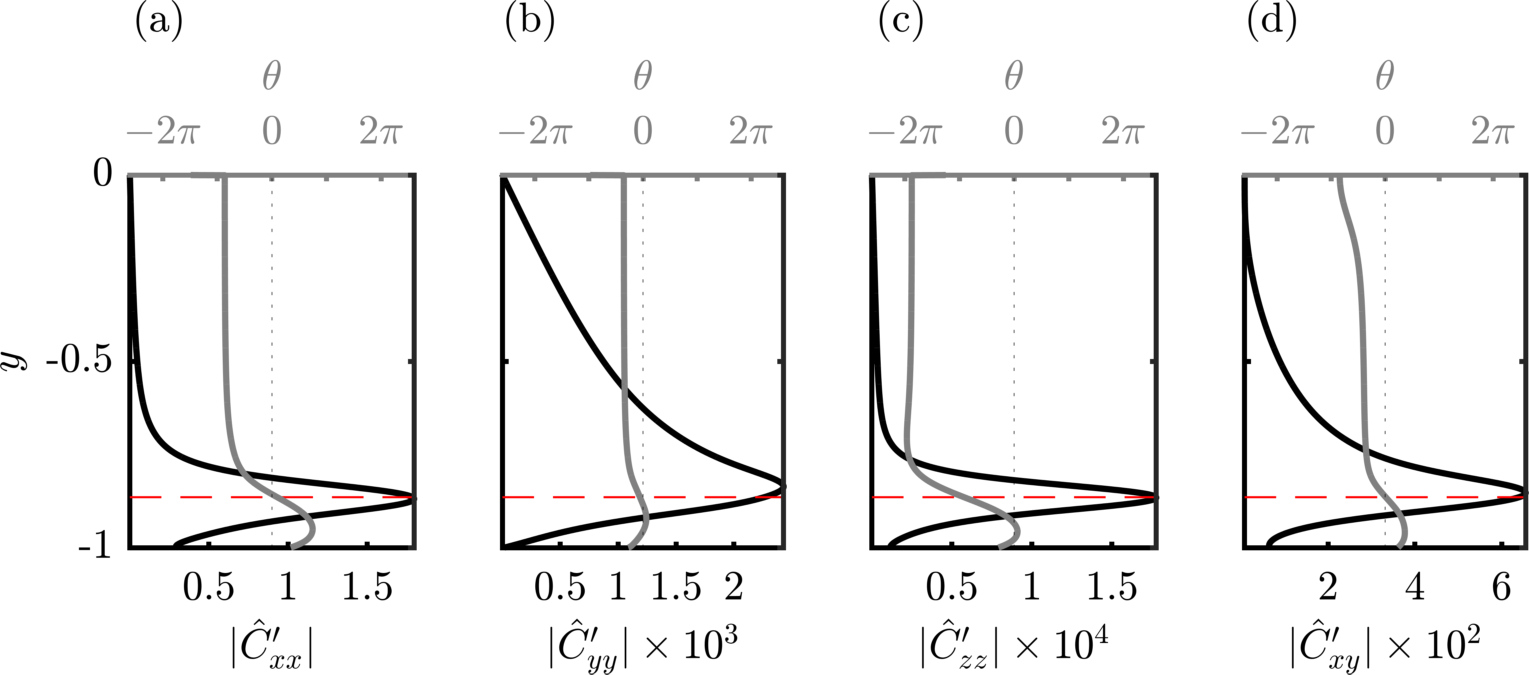}	 
		
    \parbox{\textwidth}{
    \hspace{0.85in} $|\mathsfi{\hat{C}}_{xx}'|$ 
  \hspace{0.7in} $|\mathsfi{\hat{C}}_{yy}'|\times 10^3$
    \hspace{0.6in} $|\mathsfi{\hat{C}}_{zz}'|\times 10^4$
        \hspace{0.55in} $|\mathsfi{\hat{C}}_{xy}'|\times 10^2$
}
    
		\vspace{0.2in}
		
		\centering
            \parbox{\textwidth}{
      \hspace{0.5in} (e)
      \hspace{1in} (f)
      \hspace{1in} (g)
      \hspace{1in} (h)
    }
  
		\hspace{0.11in}\includegraphics[scale=1.0,trim={0.in 0.2in 0in 0.2in},clip]{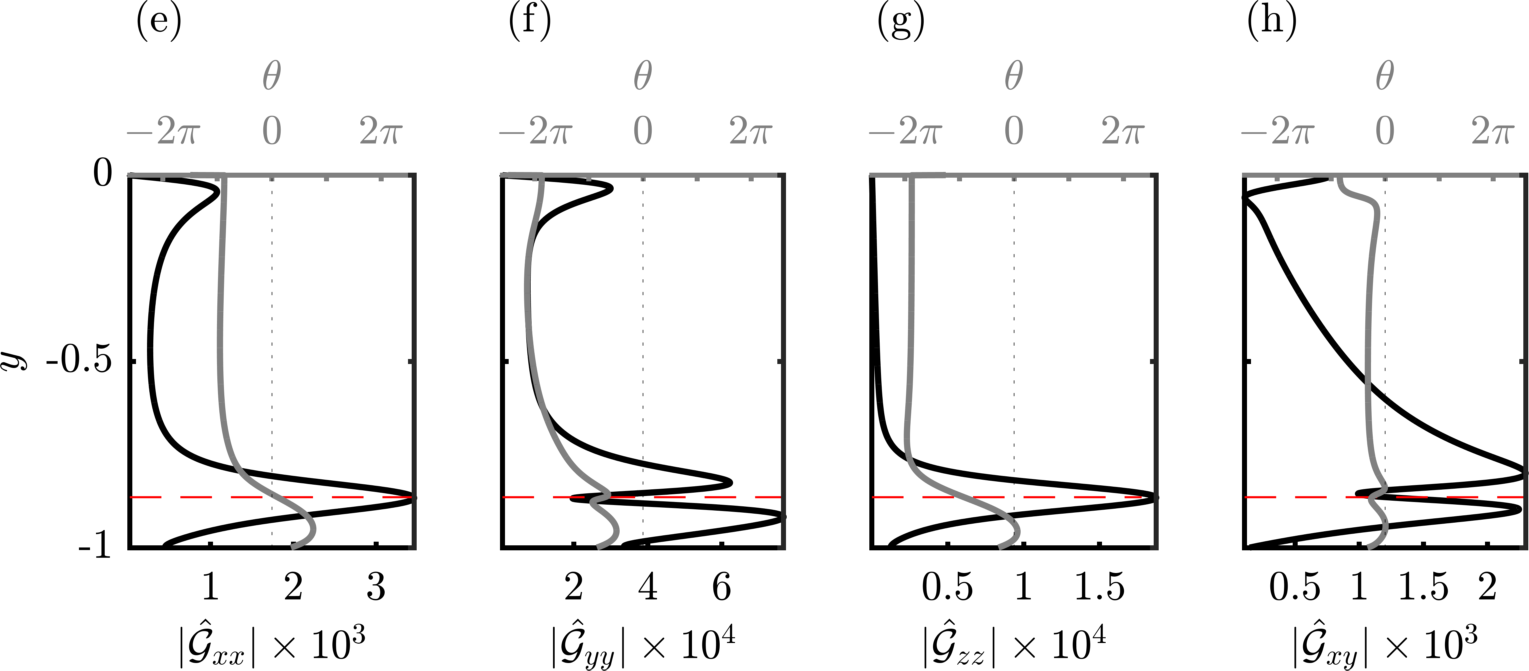}
    
        \parbox{\textwidth}{
      \hspace{0.75in} $|\mathcal{\hat{G}}_{xx}|\times 10^3$ 
      \hspace{0.55in} $|\mathcal{\hat{G}}_{yy}|\times 10^4$
      \hspace{0.55in} $|\mathcal{\hat{G}}_{zz}|\times 10^4$
      \hspace{0.525in} $|\mathcal{\hat{G}}_{xy}|\times 10^3$
    }	
	}
	
	\caption{$\Wie=6.67$. Components of the initial perturbation tensor Fourier mode:  (a)--(d) in the native form, $\hat{\mathsfbi{C}}'|_{t=0}$, and (e)--(h) in the form of a tangent on $\Pos_3$, $\bld{\hat{\mathcal{G}}}|_{t=0}$. In all panels: solid black lines are the absolute magnitudes of the modes, solid grey lines are the phase angles $\theta$. The horizontal red dashed line is the location of the critical layer, and the vertical thin black dotted line marks zero phase angle. Note: As indicated in the axes labels, the perturbation values are normalized to optimize the clarity of the plots.}
	
	\label{fig:Ceigenmode_2}
\end{figure}

The initial condition is constructed from the superposition of the base state and the instability mode, 
\begin{align}
	\bld{u}|_{t=0}  &= \bld{\overline{u}} + \Real\{\bld{\hat{u}}'|_{t=0}\e{\text{i}k_x  x}\}, \qquad
	\mathsfbi{C}|_{t=0}  = \mathsfbi{\overline{C}} + \Real\{\mathsfbi{\hat{C}}'|_{t=0}\e{\text{i} k_x x}\}, \label{IC}
\end{align}
where $(\bld{\overline{u}},\mathsfbi{\overline{C}})$ is the laminar flow, $(\bld{u}',\mathsfbi{C}')$ is the (additive) perturbation, and hats denote quantities that are Fourier transformed in the $x$ direction\textemdash only the $k_x=1$ Fourier component is non-zero for the initial perturbation.
The initial perturbation magnitude is fixed at $0.01\%$ of the bulk velocity.
The non-zero components of $\mathsfbi{\hat{C}}'|_{t=0}$  are shown in figures \ref{fig:Ceigenmode_1} and \ref{fig:Ceigenmode_2}.
The figures also show the non-zero components of  $\bld{\hat{\mathcal{G}}}|_{t=0}$ which is evaluated by pre- and post-multiplying the second equation in (\ref{IC}) with $\mathsfbi{\overline{F}}^{-1}=\mathsfbi{\overline{F}}^{-\mathsf{T}}$, and is the perturbation tangent along $\Pos_3$,
\begin{align}
\mathsfbi{G}|_{t=0} = \mathsfbi{I} + \Real\{\bld{\hat{\mathcal{G}}}|_{t=0}\e{\text{i} k_x x}\}.
\end{align}
The correct form of the tangent on $\Pos_3$, $\bld{\hat{\mathcal{G}}}|_{t=0}$ reveals details about the perturbation that are not apparent from $\mathsfbi{\hat{C}}'|_{t=0}$.
We describe the most salient of these points below.

We first consider $W1.83$.
The perturbation streamwise normal stretch $\hat{\mathsfi{C}}'_{xx}|_{t=0}$ is shown in figure \ref{fig:Ceigenmode_1}(a), and suggests that the polymer perturbation rapidly tapers off above the critical layer (the location where the instability phase speed equals the local mean velocity). 
Figure \ref{fig:Ceigenmode_1}(e), however, shows that $|(\hat{\mathcal{G}}_{xx}|_{t=0})|$ remains relatively constant from the critical layer up to 0.84 channel half-heights away from the wall. 
Since $\tr \bld{\hat{\mathcal{G}}}|_{t=0} \approx \hat{\mathcal{G}}_{xx}|_{t=0}$, the large values of $\hat{\mathcal{G}}_{xx}|_{t=0}$ imply that the change in the volume-ratio induced by the perturbation, relative to the mean, is similar deep in the channel and at the critical layer.
This is a reflection of the fact that the mean volume is smaller closer to the centreline and therefore deformations with respect to it appear, in general, larger than those with respect to the near wall configuration.
This accurate account of the perturbation magnitude is only feasible by examining it in the correct form along the tangent to $\Pos_3$, i.e.\,$\bld{\mathcal{G}}$, and also highlights the drawback of an additive, or vector space, view.

While the maximum $|(\hat{\mathsfi{C}}'_{yy}|_{t=0})|$ is located at the critical layer, the largest $y$-direction normal stretch actually occurs at the wall as shown by $|(\hat{\mathcal{G}}_{yy}|_{t=0})|$.
Thus, examining the perturbation in its correct form, $\bld{\mathcal{G}}$, dispels possible confusion regarding the dominance of the perturbation at the critical layer. 
The component $\hat{\mathcal{G}}_{xy}|_{t=0}$ captures the shearing deformation induced by perturbation on the mean configuration (see the discussion  in \S \ref{sec:wnd}).
A peak in $|(\hat{\mathcal{G}}_{xy}|_{t=0})|$ appears below the critical layer, which demonstrate that the perturbation induces the most shearing of the mean configuration at that location.  
This effect is missing from $|(\hat{\mathsfi{C}}'_{xy}|_{t=0})|$.

Unlike $W1.83$, at the higher Weissenberg number ($W6.67$), figures \ref{fig:Ceigenmode_2}(a)--(e) show that both $\hat{\mathsfi{C}}'_{xx}|_{t=0}$ and  $|(\hat{\mathcal{G}}_{xx}|_{t=0})|$ taper off above the critical layer.
In this case, the peak $|(\hat{\mathcal{G}}_{xx}|_{t=0})|$ is no longer at the wall, but at the critical layer.    
On the other hand, while the maximum of $|(\hat{\mathsfi{C}}'_{yy}|_{t=0})|$ is located at the critical layer, the maximum of $|(\hat{\mathcal{G}}_{yy}|_{t=0})|$ is slightly below.

In the classical approach, $\mathsfbi{C}'$ does not explain the changes in the volume-ratio of the perturbed to the base conformation or the shearing deformation, because the principal axes of $\mathsfbi{C}'$ and $\mathsfbi{\overline{C}}$ are not necessarily aligned.
Thus, the geometry and relevant interpretations laid out in this paper must be kept in mind when studying perturbations to the conformation tensor.

\begin{figure}
   \centering
   \includegraphics[scale=0.9]{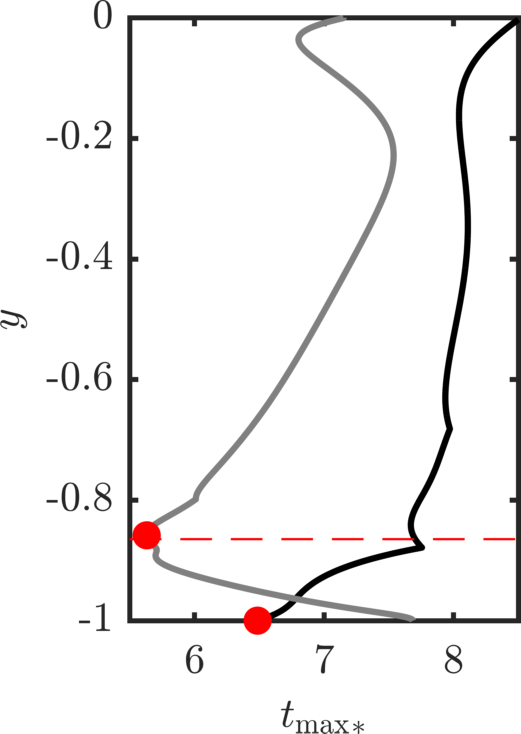}	
   \caption{Upper bound, for each wall-normal plane, on the time for which evolution of $\mathsfbi{G}$ along Euclidean lines remains positive--definite, as defined in (\ref{TPosCon}). The time is normalized by the growth rate, ${t_{\max}}_* = \omega_{\textrm{i}} t_{\max}$.  Black line is $W1.83$ and grey line is $W6.67$.  The solid red circle markers (\redfilledcirc) indicate the lowest ${t_{\max}}_*$ for each curve: ${t_{\max}}_*=6.48$ for $\Wie=1.83$, and  ${t_{\max}}_*=5.63$ for $\Wie=6.67$. The location of the critical layer for $W6.67$, which is approximately the same for $W1.83$, is shown as a red dashed line. }
   \label{fig:Tmax}
 \end{figure}

The longest time, $t_{\max}$, for which the conformation tensor can evolve along a Euclidean path on $\Pos_3$ was derived (equation \ref{TPosCon}). 
This time was evaluated from the initial condition at each wall-normal plane in the channel, and is shown in figure \ref{fig:Tmax}.
The minimum value of $t_{\max}$ represents the upper bound on the duration of linear evolution for the entire domain.
As per \eqref{TPosCon}, choosing a larger initial perturbation magnitude $\varepsilon$ decreases $t_{\max}$.
For $W1.83$ the minimum $t_{\max}$ is located at the wall, while for $W6.67$ it is located at the critical layer, which suggests that the critical layer plays an important role in the latter case.
This result is consistent with figures \ref{fig:Ceigenmode_1} and \ref{fig:Ceigenmode_2}: 
The dominant perturbation for $W1.83$ is at the wall, while that for $W6.67$ is closer to the critical layer.

\subsection{Time evolution of instability waves}

\label{sec:TS:time}

In order to track the nonlinear temporal evolution of the unstable modes, we consider the following scalar quantities,
\begin{align}
E   \equiv  \frac{1}{L_{x}L_{y}L_{z}}\int|\bld{u}'|^2 \,\text{d}x\,\text{d}y\,\text{d}z,  \quad
J   \equiv \frac{1}{L_{x}L_{y}L_{z}}\int d^2(\mathsfbi{\overline{C}},\mathsfbi{C})  \,\text{d}x\,\text{d}y\,\text{d}z  \label{EJDefn}
\end{align}
where $d^2(\mathsfbi{\overline{C}},\mathsfbi{C})=d^2(\mathsfbi{I},\mathsfbi{G}) =\tr\bld{\mathcal{G}}^2$, and $\bld{\mathcal{G}}=\log \mathsfbi{G}$.  
The quantity $E$ is the volume-averaged perturbation kinetic energy, or the Euclidean norm associated with $\bld{u}'$, and was used by \citet{Lee2017} to characterize viscoelastic Tollmien--Schlichting waves. 
While their fluctuation was defined with respect to the instantaneous mean flow, here the reference laminar state is used.
We propose the quantity $J$ to evaluate the evolution of the polymer deformation, which is the volume-averaged squared geodesic distance of $\mathsfbi{C}$ away from $\mathsfbi{\overline{C}}$ \citep{Hameduddin2018a}.
The geodesic distance is the natural way to measure the size of the fluctuating deformation in $\mathsfbi{C}$ because we cannot define a norm on $\Pos_3$ due to the lack of a vector space structure.
In the classical approach, one would perhaps use the Frobenius norm of $\mathsfbi{C}'$ to quantify the fluctuating polymer deformation, which would lead to a wide variety of difficulties, e.g.\,the Frobenius norm would predict that regions of negative  $\mathsfbi{C}'$ are equivalent to regions of positive $\mathsfbi{C}'$. 
However, this is manifestly not the case (c.f. discussion in \,\S\ref{sec:geometry}): regions of positive (negative) $\mathsfbi{C}'$ represent polymer expansion (compression) and while expansions may be arbitrarily large, compressions cannot be. 
The difficulties cited above arise when Euclidean concepts are foisted upon a non-Euclidean manifold and can be completely eliminated by adopting the mathematically consistent viewpoint that treats $\Pos_3$ as a Riemannian manifold.
The quantity $J$ is proposed in such spirit.

\begin{figure}
	\centering
	\includegraphics[width=0.85\textwidth]{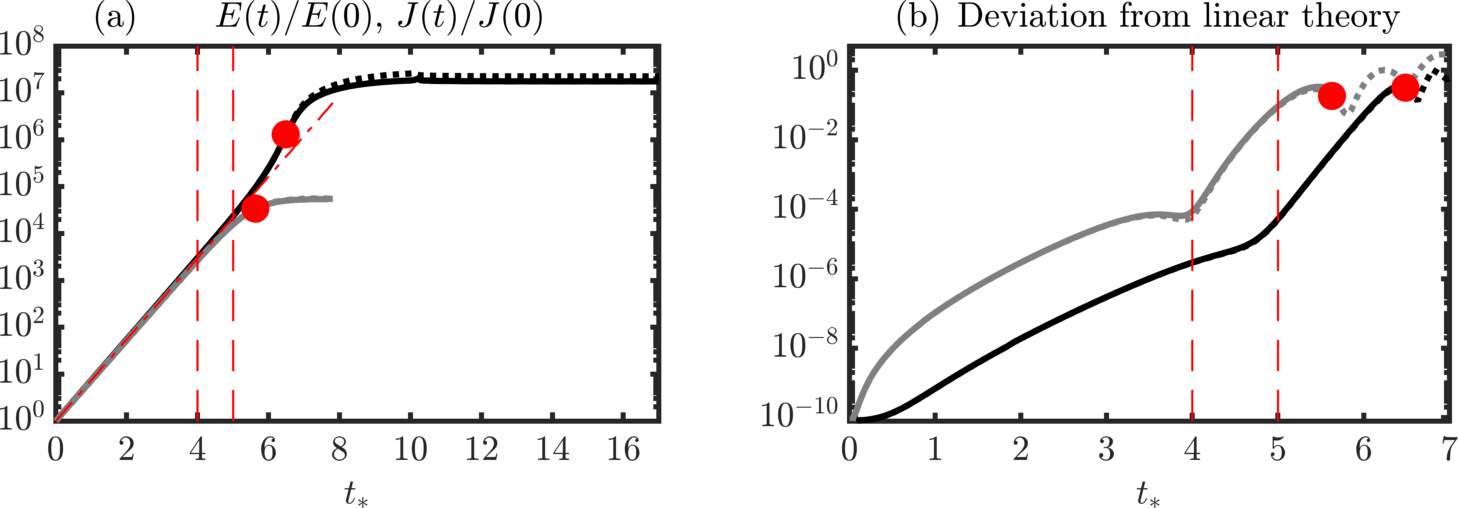}	
	\caption{ (a) Evolution of $E$ and $J$, as defined in (\ref{EJDefn}), normalized by the initial values.
	For $W1.83$, $E(0)= 2.05\times 10^{-9}$ and $J(0)= 1.53\times 10^{-7}$;  for $W6.67$, $E(0)= 1.97\times 10^{-9}$ and $J(0)= 2.03\times 10^{-6}$. 
	Solid lines (\linesolids, {\color{gray}\linesolids}) are $J(t)/J(0)$ while dotted lines (\linedotted, {\color{gray}\linedotted}) are $E(t)/E(0)$.   The red dash-dot line ({\color{red}\linedshdot}) is an extrapolation based on the growth rate predicted by linear theory. (b) Deviation from linear theory; solid lines (\linesolids, {\color{gray}\linesolids}) are the deviations defined by (\ref{geodesic_deviation}) and dotted lines (\linedotted, {\color{gray}\linedotted})  are the deviations defined by (\ref{geodesic_deviation_true}).  
    In both (a) and (b): black lines are $W1.83$ , grey lines are $W6.67$, thin dashed red lines ({\color{red} \linedashed})  mark $t_*=4$ and $t_*=5$, and the solid red circle markers (\redfilledcirc) indicate the minimum ${t_{\max}}_*= t_{\max} \omega_{\textrm{i}}$ as defined in (\ref{TPosCon}) and also marked in figure \ref{fig:Tmax} (${t_{\max}}_*=6.48$ for $W1.83$ and ${t_{\max}}_*=5.63$ for $W6.67$). }
	\label{fig:EJ}
\end{figure}

The time evolutions of $E$ and $J$ for both cases are shown in figure \ref{fig:EJ}(a) as a function of the normalized time $t_*=\omega_{\text{i}}t$.
For $W1.83$, the evolution of $E$ matches the prediction by linear theory for $t_* \lesssim 5$, then shows super-exponential growth and finally saturates at $t_* \approx 10$.
For $W6.67$, the evolution agrees with linear theory for $t_* \lesssim 4$, shows no super-exponential growth and saturates at $t_*\approx 8$.
The different behaviours when the curves deviate from linear theory suggest that different physical mechanisms are at play at the two Weissenberg numbers.
The evolutions of normalized $J$ closely match those of normalized $E$.
They are also consistent with an assumption that $\mathsfbi{G}$ initially evolves along a linear approximation of the geodesic on $\Pos_3$ emanating from $\mathsfbi{I}$ because for $t_* \lesssim 5$,
\begin{align}
d^2(\mathsfbi{\overline{C}},\mathsfbi{C})
=  \text{tr}\,\log^2\mathsfbi{G} 
&\approx  \text{tr}\, \log^2\left(\mathsfbi{I}  + \bld{\mathcal{G}}   \right)   \approx  \text{tr}\,   \bld{\mathcal{G}}^2 \sim \varepsilon^2 \e{2\omega_{\textrm{i}} t },  \label{EuclideanPath}
\end{align}
where we used the matrix Mercator series \citep{Higham2008}, and assumed  $\|\bld{\mathcal{G}}\| \sim \varepsilon \ll 1$.

As the Tollmien--Schlichting waves evolve in time, nonlinear effects become appreciable.
For the velocity field, the role of nonlinearity can be quantified using the Euclidean norm of the deviation of the velocity field from the prediction by linear theory. 
The question then becomes: how do we quantify the deviation of the conformation tensor obtained using direct numerical simulations from the prediction by linear theory?
In the spirit of geometric consistency, we evaluate the squared geodesic distance between the two,
 \begin{align}
\frac{1}{L_{x}L_{y}L_{z}}\int d^2(\mathsfbi{G},\mathsfbi{I} + \text{Re}\{\bld{\hat{\mathcal{G}}}|_{t=0}\e{\text{i} (k_x x + \omega t)}\}) \,\text{d}x\,\text{d}y\,\text{d}z, \label{geodesic_deviation}
\end{align} 
which is plotted in figure \ref{fig:EJ}(b).
The deviation in \eqref{geodesic_deviation} is a measure of the importance of nonlinear effects.
Predictably, the deviation from linear theory is only valid as long as the linear approximation $\mathsfbi{I} + \text{Re}\{\bld{\hat{\mathcal{G}}}|_{t=0}\e{\text{i} (k_x x + \omega t)}\}$ remains positive--definite, and hence $t < t_{\max}$.
The upper bounds on that time, calculated using (\ref{TPosCon}), were shown in figure \ref{fig:Tmax} and are also indicated as solid red circle markers in figure \ref{fig:EJ}.
For longer times, we can instead evaluate
\begin{align}
	\frac{1}{L_{x}L_{y}L_{z}}\int d^2(\mathsfbi{G}, \e{\text{Re}\{\bld{\hat{\mathcal{G}}}|_{t=0}\e{\text{i} (k_x x + \omega t)}\} }) \,\text{d}x\,\text{d}y\,\text{d}z. \label{geodesic_deviation_true}
\end{align}
The quantity \eqref{geodesic_deviation_true} is also shown in figure \ref{fig:EJ}.
The evolution of \eqref{geodesic_deviation} is virtually indistinguishable from \eqref{geodesic_deviation_true} because the difference between them is $\mathcal{O}(\varepsilon^2\e{\varepsilon})$. 
As illustrated in the figure, however, the latter can be extended indefinitely since we are evaluating the deviation away from a perturbation along a geodesic, which is guaranteed to remain on the manifold.

For $W1.83$, the deviation between DNS and linear theory increases at a relatively slow rate up to $t_*\approx 5$, when it begins to grow faster until $t_*\approx 6.5$. 
The initial growth in the deviation is associated with part of the region matching linear theory in the evolution of $J$, $0 \lesssim t_* \lesssim 5$, while the later growth may be associated with super-exponential growth in $E$ and $J$ that appears before saturation.
As discussed earlier, at even longer times than $t_*\approx 6.5$, the linear approximation does not remain positive--definite everywhere in the domain and thus the deviation cannot be calculated further.
For $W6.67$, we see a similar slow initial growth in the deviation from linear theory until $t_* \approx 4$, where the growth abruptly becomes faster.

The above discussion emphasizes the importance and value of the geometry of $\Pos_3$:  
It allowed us to formulate measures of the deviation away from linear theory in \eqref{geodesic_deviation} and \eqref{geodesic_deviation_true}.  
It also enabled the interpretation of a perturbation as an excursion along a geodesic, and furnished us with a measure of the deviation \eqref{geodesic_deviation_true} that remains valid at large times.

\begin{figure}
  \centering
  
  \parbox{\textwidth}{\hspace{0.25in} (a)}
  
  \includegraphics[width=0.85\textwidth,trim={0.000in 0.240in 000in 0.000in},clip]{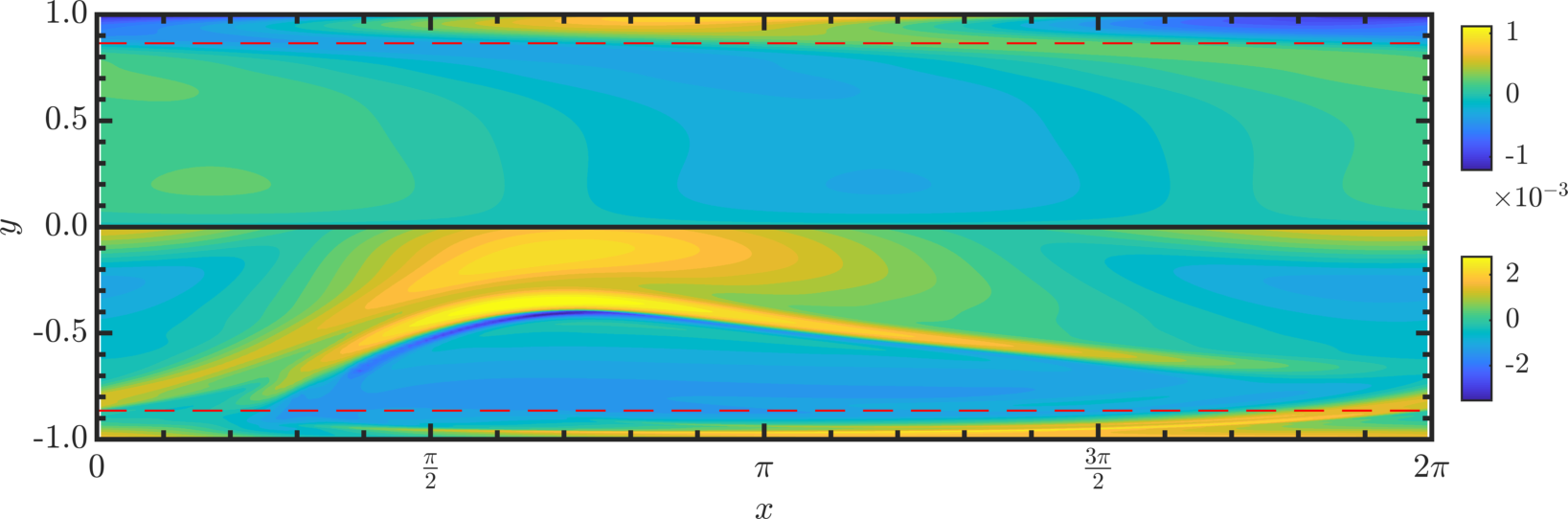}
  
  \vspace*{0.1in}
  
    \parbox{\textwidth}{\hspace{0.25in} (b) }
  
\hspace{-0.075in}    \includegraphics[width=0.85\textwidth,trim={0.000in 0.000in 000in 0.000in},clip]{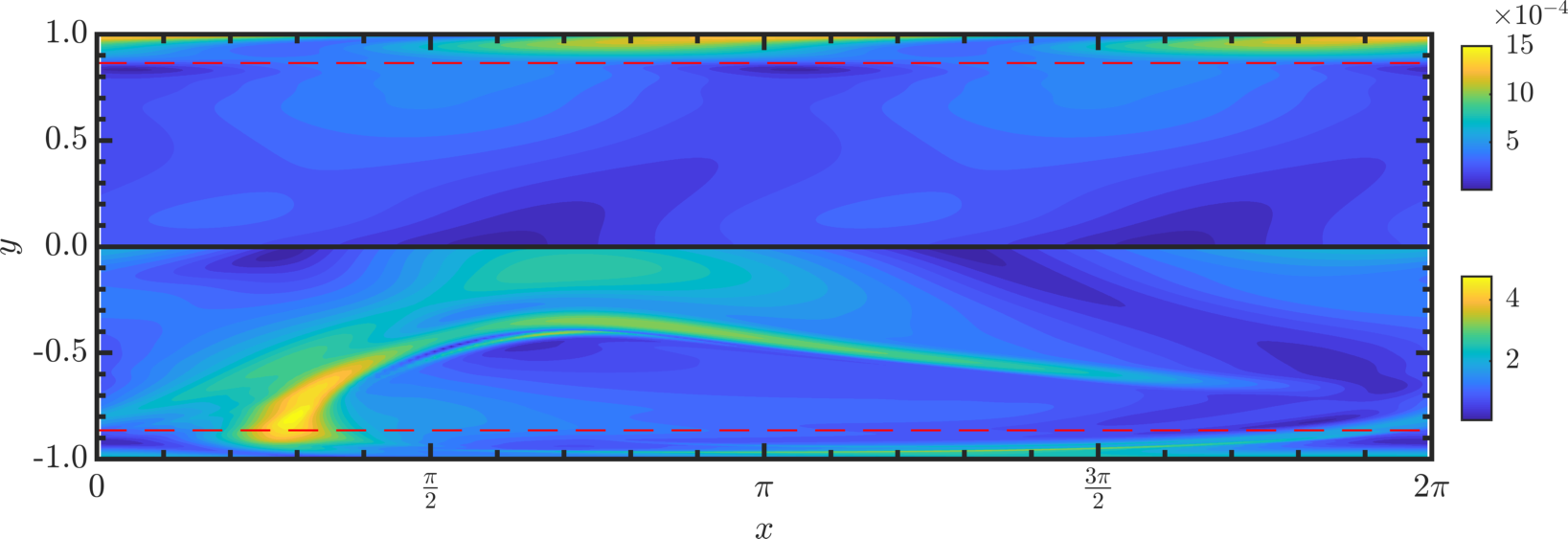}

 	\caption{Case $W1.83$ at $t_*\in\{0,17\}$. Isocontours of (a) logarithmic volume ratio, $\tr \bld{\mathcal{G}}$, (b) geodesic distance from the laminar, $\sqrt{\tr \bld{\mathcal{G}}^2}$. 
   In both panels, the top half ($y\in(0,1]$) of the channel is at $t_*=0$ and the bottom half ($y\in[-1,0)$) of the channel is at $t_*=17$.
   The solid black line is the channel centreline and the dashed red lines are the locations of the critical layers.} 
 
  	\label{fig:vol_geodesic_1}
\end{figure}

\begin{figure}
  \centering
  
  \parbox{\textwidth}{\hspace{0.25in} (a)}
  
  \includegraphics[width=0.85\textwidth,trim={0.000in 0.240in 000in 0.000in},clip]{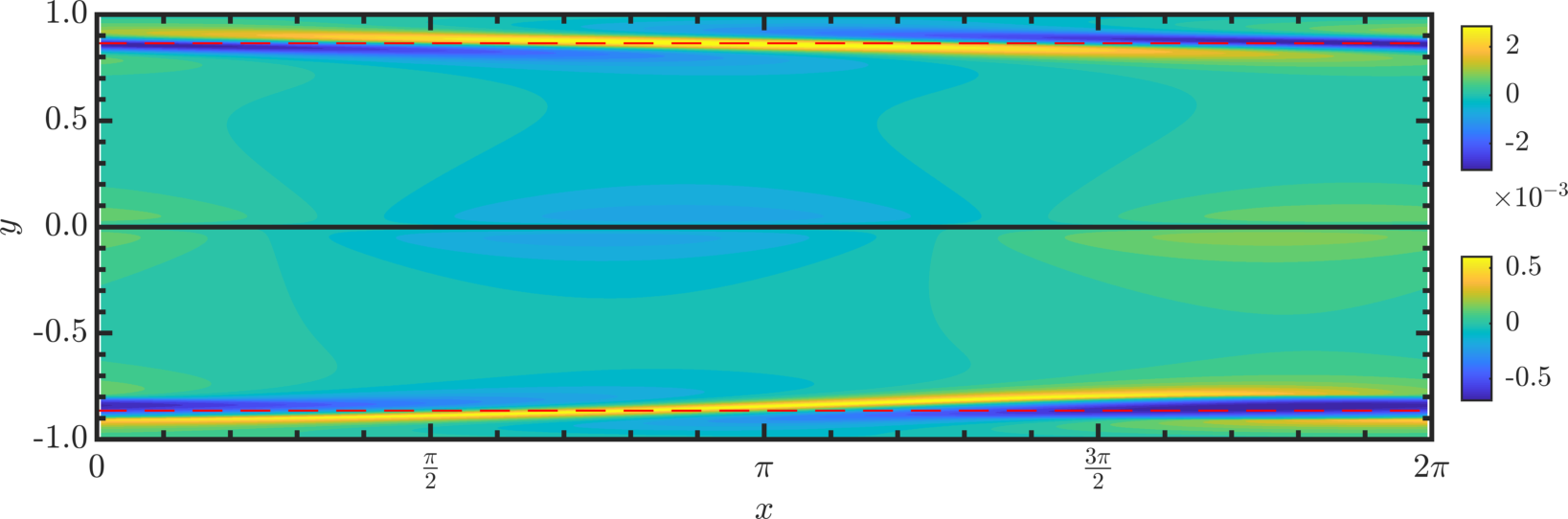}
  
  \vspace*{0.1in}
  
  \parbox{\textwidth}{\hspace{0.25in} (b) }
  
  \hspace{-0.075in}    \includegraphics[width=0.85\textwidth,trim={0.000in 0.000in 000in 0.000in},clip]{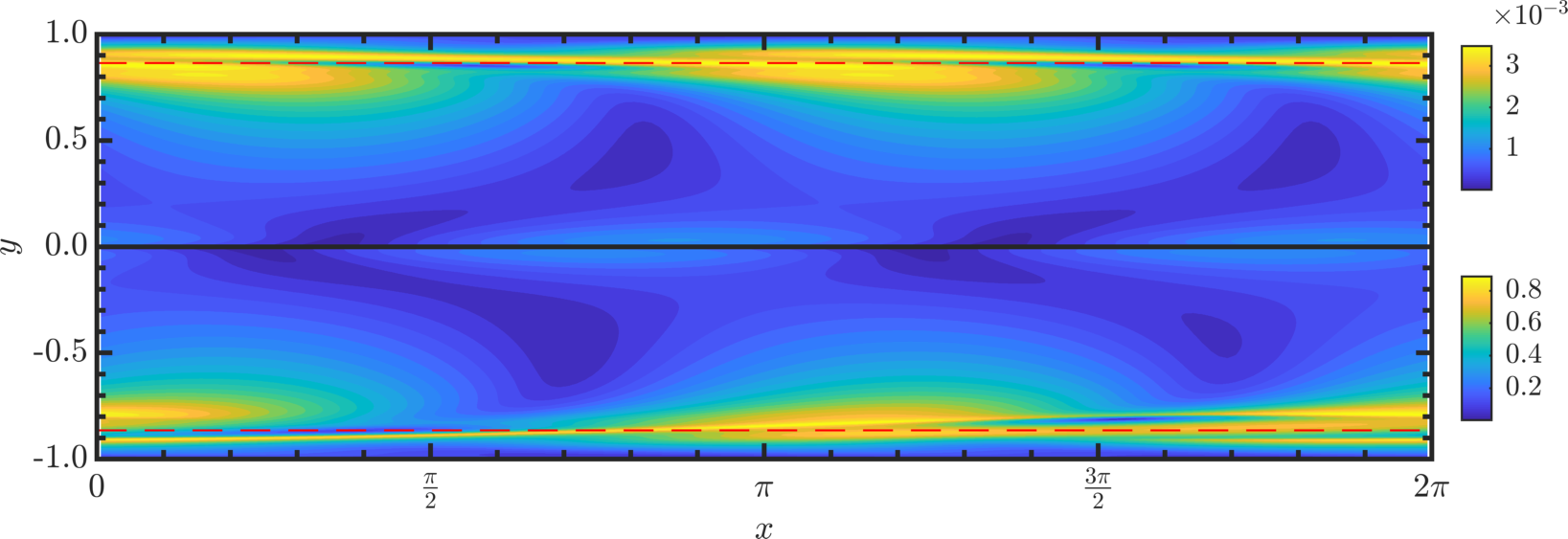}
  
   	\caption{Case $W6.67$ at $t_*\in\{0,8\}$. Isocontours of (a) normalized logarithmic volume ratio, $\tr \bld{\mathcal{G}}$, (b) normalized geodesic distance from the laminar, $\sqrt{\tr \bld{\mathcal{G}}^2}$. 
      In both panels, the top half ($y\in(0,1]$) of the channel is at $t_*=0$ and the bottom half ($y\in[-1,0)$) of the channel is at $t_*=8$.
      The solid black line is the channel centreline and the dashed red lines are the locations of the critical layers.} 
    
     	\label{fig:vol_geodesic_2}
\end{figure}

Nonlinear effects eventually lead to the saturated state shown in figure \ref{fig:EJ}. 
It is instructive to compare the differences in polymer deformation between the initial linear stage and the saturated condition.
We quantify this difference using the logarithmic volume ratio \eqref{logVol}, here abbreviated as LVR, and the geodesic distance from the laminar base flow \eqref{geodesicDefn}, which we will refer to here simply as the geodesic distance.
The LVR is the ratio of the volume of $\mathsfbi{C}$ to the volume of $\mathsfbi{\overline{C}}$ and describes whether a deformation is volumetrically expansive or compressive, and by how much relative to the reference state.  
The geodesic distance is analogous to the norm of the velocity, but for the conformation tensor.
These two measures together allow for a succinct, yet rich picture of the deformation.
For example, a volume-preserving purely shearing deformation can be detected since it will lead to zero LVR but nonzero geodesic distance.

The isocontours of the LVR for $W1.83$ differ significantly between the initial condition and the saturated state (figure \ref{fig:vol_geodesic_1}(a)).
For the initial condition, the most significant variations occur near the wall below the critical layer, with only very weak volume-ratio changes elsewhere.
On the other hand, the main activity in the saturated state is focused away from the wall: 
A thin, elongated region of large volume-ratio expansion is centred at $y\approx-0.5$, and large region of volume-ratio expansion is near $y \approx -0.1$ and is connected to the wall via a filament of expansive stretch.
The geodesic distance shown in figure \ref{fig:vol_geodesic_1}(b) is consistent with these volumetric changes.
However, the geodesic distance peaks at $x\approx 0.3\pi$ in the critical layer, even though this region only shows relatively small LVR in figure \ref{fig:vol_geodesic_1}(a).
These results imply that the polymer is undergoing a shearing deformation at the this location.
The significant differences between the flow structures in the linear and saturated stages demonstrate that the mechanism for the initial growth of the Tollmien--Schlichting wave is disrupted in the latter stage. 
Only at saturation do we observe large volume-ratio changes near the centreline, and the localized largely shearing deformation near the critical layer.

The isocontours of the LVR for $W6.67$, shown in figure \ref{fig:vol_geodesic_2}(a), are strikingly similar at the initial and saturated states.
The main volume-ratio change is near the critical layer, and remains so in the saturated state.
The geodesic distance in  \ref{fig:vol_geodesic_2}(b) confirms this finding, with little shearing deformation evident.
Both measures also show that there is a spreading of the stretched/compressed region from the critical layer towards the centreline.
Both the initial condition and the saturated state are dominated by the $k_x=1$ mode, suggesting that the nonlinear terms in the governing equations do not significantly alter the perturbation spectra.

\begin{figure}
  \centering
  
  \parbox{\textwidth}{\hspace{0.25in} (a)}
  
\includegraphics[width=0.85\textwidth,trim={0.000in 0.240in 000in 0.000in},clip]{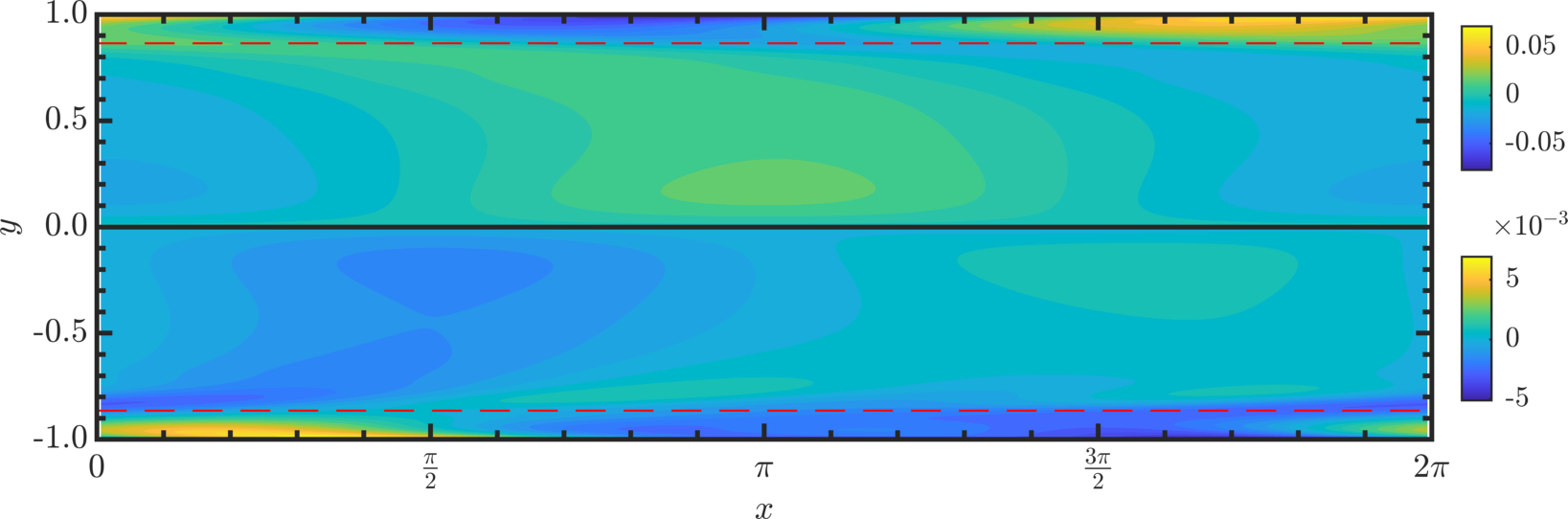}

\vspace*{0.1in}

\parbox{\textwidth}{\hspace{0.25in} (b)}

\includegraphics[width=0.85\textwidth,trim={0.000in 0.240in 000in 0.000in},clip]{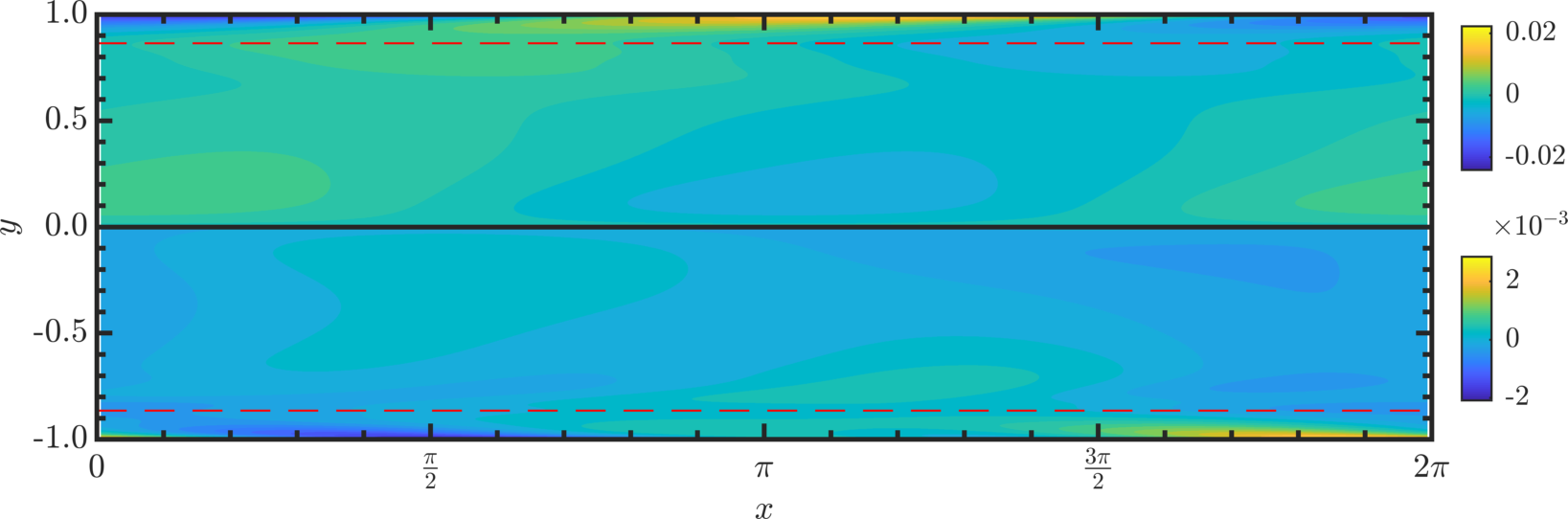}

\vspace*{0.1in}

\parbox{\textwidth}{\hspace{0.25in} (c)}

\includegraphics[width=0.85\textwidth,trim={0.000in 0.000in 000in 0.000in},clip]{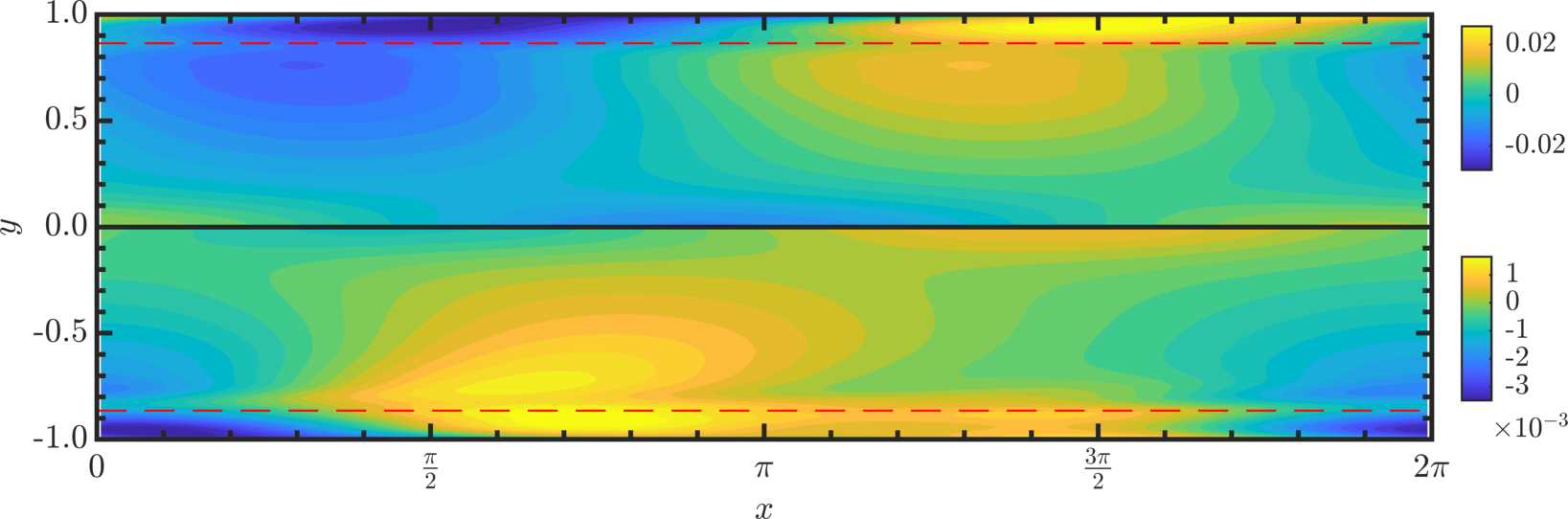}

\caption{Components of $\varepsilon^k\bld{\mathcal{G}}_k$ with $k\in\{1,2\}$ for $W1.83$ at $t_*=4$, i.e. the tangents on $\Pos_3$ that appear in the weakly nonlinear deformation expansion (\ref{nl_expansion1}).
  (a) $\varepsilon^k\mathcal{G}_{k_{xx}}$.
  (b) $\varepsilon^k\mathcal{G}_{k_{yy}}$.
  (c) $\varepsilon^k\mathcal{G}_{k_{xy}}$.
In all panels, the top half ($y\in(0,1]$) of the channel is $k=1$ and the bottom half ($y\in[-1,0)$) of the channel is $k=2$.
 The solid black line is the channel centreline and the dashed red lines are the locations of the critical layers.
}

\label{fig:wnd_W1p83}
\end{figure}

 \begin{figure}
   \centering
   
   \parbox{\textwidth}{\hspace{0.25in} (a)}
   
   \includegraphics[width=0.85\textwidth,trim={0.000in 0.240in 000in 0.000in},clip]{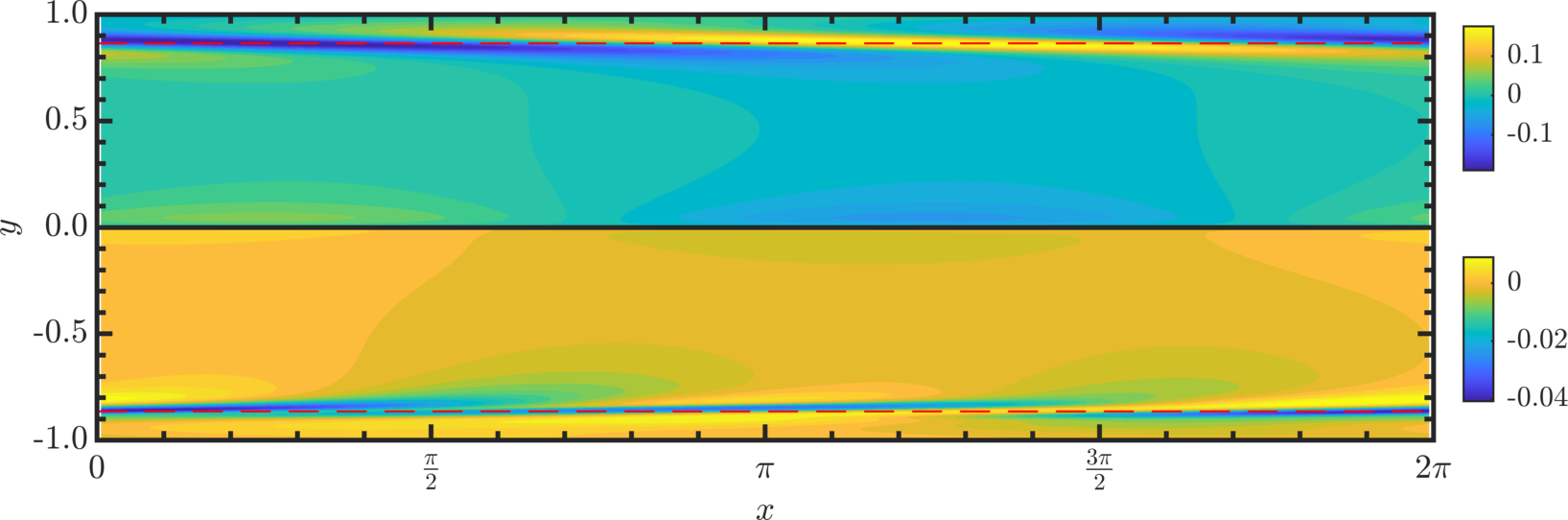}
   
   \vspace*{0.1in}
   
   \parbox{\textwidth}{\hspace{0.25in} (b)}
   
   \includegraphics[width=0.85\textwidth,trim={0.000in 0.240in 000in 0.000in},clip]{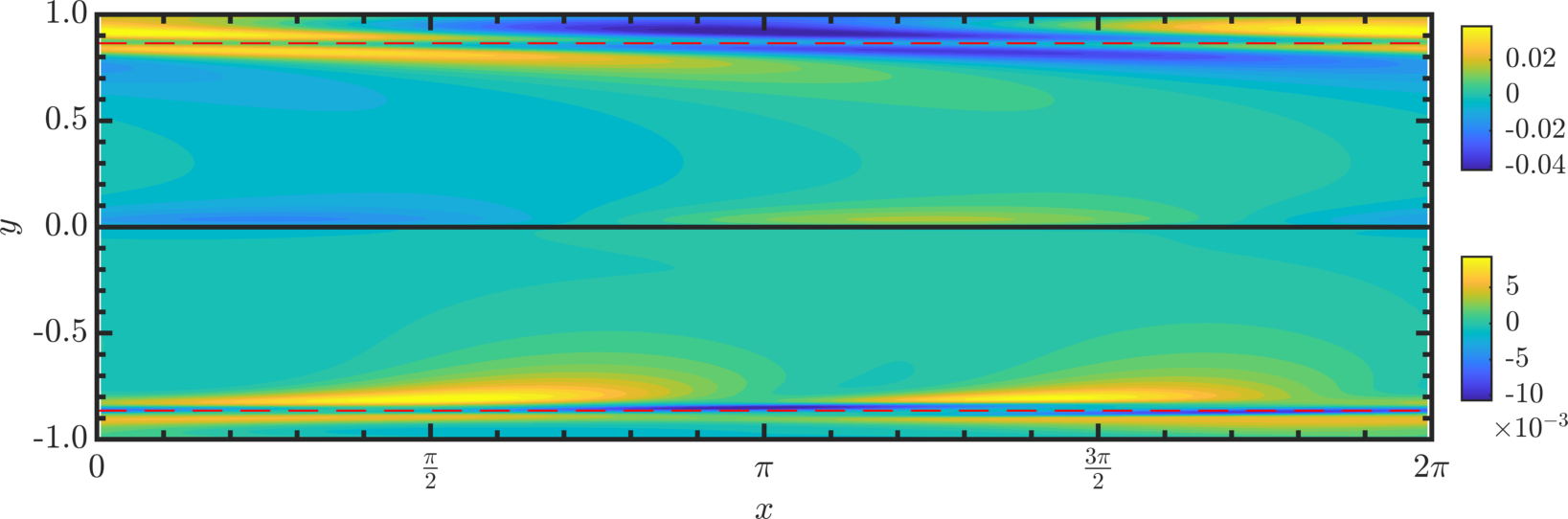}
   
   \vspace*{0.1in}
   
   \parbox{\textwidth}{\hspace{0.25in} (c)}
   
   \includegraphics[width=0.85\textwidth,trim={0.000in 0.000in 000in 0.000in},clip]{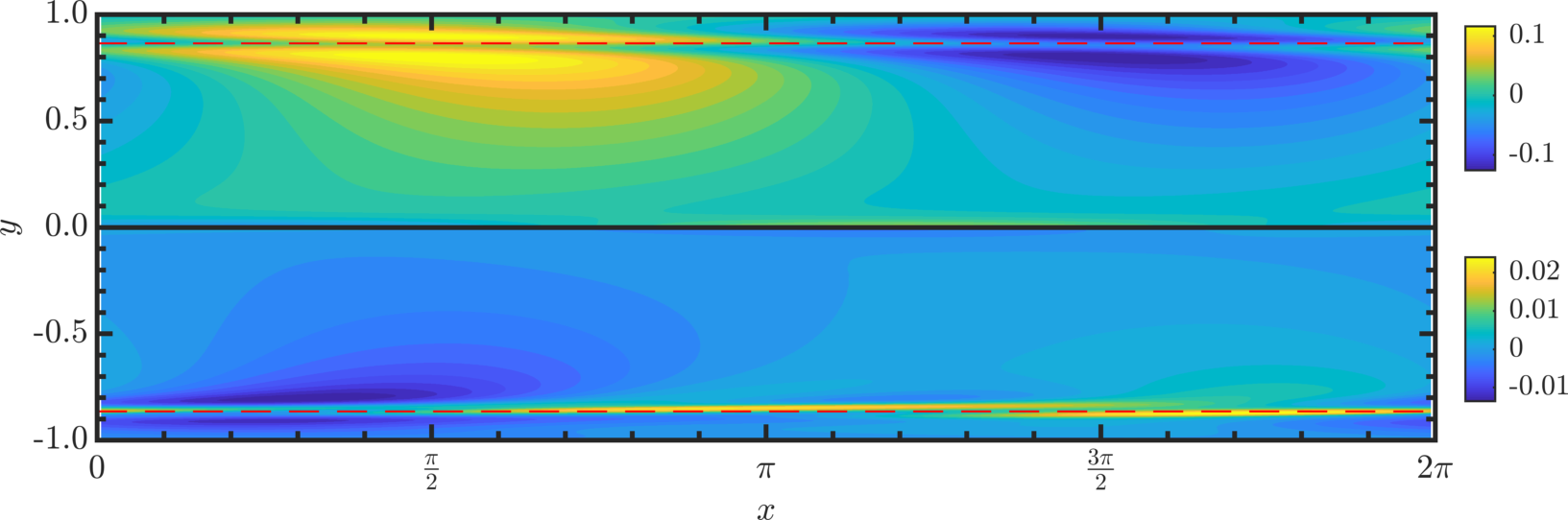}
   
   \caption{Components of $\varepsilon^k\bld{\mathcal{G}}_k$ with $k\in\{1,2\}$ for $W6.67$ at $t_*=4$, i.e. the tangents on $\Pos_3$ that appear in the weakly nonlinear deformation expansion (\ref{nl_expansion1}).
     (a) $\varepsilon^k\mathcal{G}_{k_{xx}}$.
     (b) $\varepsilon^k\mathcal{G}_{k_{yy}}$.
     (c) $\varepsilon^k\mathcal{G}_{k_{xy}}$.
     In all panels, the top half ($y\in(0,1]$) of the channel is $k=1$ and the bottom half ($y\in[-1,0)$) of the channel is $k=2$.
     The solid black line is the channel centreline and the dashed red lines are the locations of the critical layers.
}
   
   \label{fig:wnd_W6p67}
 \end{figure}

\subsection{Weakly nonlinear deformation}

\label{sec:TS:wnd}

In order to examine the initial deviation from linear theory, we adapt the approach that was used by \cite{Benney1960} to study secondary instabilities in Newtonian parallel shear flows.
For sufficiently small deviations from linear evolution, we assume that $\mathsfbi{G}$ can be expressed as a polynomial expansion, here as the weakly nonlinear deformation in (\ref{nl_expansion1}), because then we can endow the terms in the expansion with physical meaning.
The growth rate and phase speed predicted by linear theory can be used to directly calculate the first variable $\bld{\mathcal{G}}_{(1)}$ in the series (\ref{nl_expansion1}),
\begin{align}
	\varepsilon\bld{\mathcal{G}}_{(1)}= \Real\{\bld{\hat{\mathcal{G}}}|_{t=0}\e{\text{i}( k_x x - \omega t)}\}
\end{align}
where $\omega$ is the eigenvalue listed in Table \ref{tab:params}.
Once $\bld{\mathcal{G}}_{(1)}$ is available, it is straightforward to rearrange (\ref{nl_expansion1}) and obtain an approximation to the second variable $\bld{\mathcal{G}}_{(2)}$, 
\begin{align}
	\varepsilon^2\bld{\mathcal{G}}_{(2)} = \mathsfbi{G} - \left( \mathsfbi{I} + \varepsilon\bld{\mathcal{G}}_{(1)} + \frac{1}{2}\varepsilon^2\bld{\mathcal{G}}_{(1)}^2\right) + \mathcal{O}(\varepsilon^3).
\end{align}

For both Weissenberg numbers, we will perform the analysis at $t_*=4$, since the deviation from linear theory at that time is still relatively small but not negligible (see figure \ref{fig:EJ}).
The $xx$, $yy$ and $xy$ components of $\varepsilon\bld{\mathcal{G}}_{(1)}$ and $\varepsilon^2\bld{\mathcal{G}}_{(2)}$ are reported in figure \ref{fig:wnd_W1p83}, for $W1.83$.
The $xx$ components of both $\bld{\mathcal{G}}_{(1)}$ and $\bld{\mathcal{G}}_{(2)}$ in figure \ref{fig:wnd_W1p83}(a) show similarly shaped isocontours, but are out of phase with each other. 
Recall from the discussion in \S\ref{sec:wnd} and equation \eqref{logdetG_wnd} that the LVR is the sum of the trace of the terms of the weakly nonlinear deformation.
For both $\varepsilon\bld{\mathcal{G}}_{(1)}$ and $\varepsilon^2\bld{\mathcal{G}}_{(2)}$, the $xx$ component contributes more to LVR than any other component.
The ability to tease out the volume-ratio change by simply examining the terms in the expansion highlights a major benefit of using a weakly nonlinear deformation expansion, rooted in physical notions.

The isocontours of the $xy$ component of $\varepsilon\bld{\mathcal{G}}_{(1)}$ and $\varepsilon^2\bld{\mathcal{G}}_{(2)}$ are shown in figure \ref{fig:wnd_W1p83}(c). 
This component represents the shearing deformation (see the discussion in \S \ref{sec:wnd}).
For the prediction by linear theory, $\varepsilon\bld{\mathcal{G}}_{(1)}$, the perturbations are most strongly shearing above the critical layer at $y\approx-0.75$, whereas for the nonlinear correction, $\varepsilon^2\bld{\mathcal{G}}_{(2)}$, the shearing is focused closer to the critical and localized in $x$.
This nonlinear correction with significant localized shear explains the observation previously noted regarding figure \ref{fig:vol_geodesic_1}: there is a region, localized in $x$ and close to the critical layer, that shows the most significant geodesic deviation but locally only weakly changes the LVR.
These two factors indicate a shearing deformation, which is consistent with a large $xy$ component of the higher-order correction, $\varepsilon^2\bld{\mathcal{G}}_{(2)}$.
The ease with which the localized shearing deformation, whose effects appear prominently in the saturated state, was deduced simply from a consideration of a slight deviation from linear theory demonstrates the effectiveness of using physically relevant polynomial expansions.

Figure \ref{fig:wnd_W6p67} shows $\varepsilon\bld{\mathcal{G}}_{(1)}$ and $\varepsilon^2\bld{\mathcal{G}}_{(2)}$ for $W6.67$.
The isocontours of both quantities similarly show that regions of expansion/compression are concentrated near the critical layer and are dominated by the $k_x=1$ Fourier mode.
This behaviour echoes the findings in figure \ref{fig:vol_geodesic_2}.
Moreover, in that figure we noted the spreading of stretched/compressed regions near the critical layer towards the centreline.
The source of this spreading is clear in figure \ref{fig:wnd_W6p67}: the regions of largest stretch and compression are located right above the critical layer.

\section{Conclusions}
\label{sec:conclusions}
 
Perturbing the conformation tensor, while maintaining physical and geometric consistency, is a more complicated proposition than perturbing a Euclidean object like the velocity field. 
In this paper, we developed methods to perturb the conformation tensor in a linear (\ref{GTrunc1}) as well as weakly nonlinear manner (\ref{nl_expansion1}) that maintain this consistency. 
Our approach provides a way to relate a perturbation to geometric behaviour on the manifold $\Pos_3$, as well as to the polymer deformation. 
The latter allowed us to study the deformation of the polymer during the nonlinear evolution of viscoelastic TS waves.

The geometric viewpoint of perturbations and the proposed weakly nonlinear deformation approach are natural and have several benefits:
(i) The set of positive-definite tensors is not a Euclidean space and, as a result, Euclidean notions of distance and size are not meaningful.  
The geometric viewpoint provides a rigorous alternative to formulate quantitative measures of the polymer deformation. 
This approach was adopted to measure the magnitude of the polymer deformation in TS waves in channel flow, using the spatially averaged geodesic distance.
(ii) The weakly nonlinear deformation expansion, as opposed to a standard polynomial expansion, has an explicit and powerful physical interpretation. 
For example, the diagonal components of each term contribute additively to the volume-ratio, and the off-diagonal components represent volume ratio-preserving deformation.
We leveraged this physical interpretation to explain how the evolution of TS waves in channel flow deviates from linear theory, e.g.\,we found that strong localized shearing appears near the critical layer for the $\Wie=1.83$ case\textemdash a prominent signature of this initial shearing appeared in the saturated state.
(iii) We can formulate a perturbation to the base-flow conformation tensor that is a geodesic emanating from the base point on the manifold, i.e.\,weakly nonlinear deformation with a single non-zero tangent. 
 While a classical polynomial expansion would not guarantee, in general, positive--definiteness for any perturbation amplitude, this geodesic is always guaranteed to remain on the manifold.
  We thus have a way of generating arbitrarily large initial perturbations.
(iv) The classical linear-theory representation was shown to be a local approximation of the geodesic.  
This new insight bridges classical linear theory with the mathematically powerful Riemannian framework which is natural for positive--definite tensors, and has physically meaningful interpretations (see ii above). 
Applying this relation to TS waves, we deduced the relative magnitudes of perturbations and determined the action on the laminar base flow.
 (v) The geometry leads to a constraint on the evolution of unstable modes: an upper bound on the maximum time  an unstable mode can evolve along Euclidean lines. This upper bound, which is determined without reference to the nonlinear dynamics, was found to be surprisingly relevant to the nonlinear evolution of TS waves: it provided a good estimate of the actual time when the nonlinear effects become significant, and correctly predicted the importance of the critical layer in the saturation process at high $\Wie$.

An interesting implication of the present work, which is not considered here, concerns linear non-modal stability analysis. 
In the present work, we showed that the appropriate inner product vis-\`{a}-vis perturbations to the conformation tensor is the one induced locally at the base conformation, by the global metric on the manifold of positive--definite tensors. 
This scalar product is not the Frobenius inner product, and thus affects the results of non-modal stability analyses which depend critically on the form of the inner product and associated induced norm.
For instance, one may seek the unit-norm initial perturbation to the conformation tensor that maximizes kinetic energy growth at a later time.
This optimal will vary depending on the norm used to constrain the initial condition and, as a result, will be different for our present formulation compared to the classical one.
In a similar vein, the present approach also provides a sensible way to measure, and thus optimize, the growth in the conformation tensor using the local geodesic distance.

\section*{Acknowledgement}
This work was sponsored in part by the National Science Foundation under grants CBET-1511937 and 1652244, and by a Johns Hopkins University Catalyst award.  
Computational resources were provided by the Maryland Advanced Research Computing Centre (MARCC).

\bibliographystyle{jfm}
\bibliography{manuscript}

\end{document}